\documentclass[
 reprint, amsmath, amssymb,
 aps, dvipsnames, nofootinbib, prd, floatfix, nofootinbib,
 nobibnotes,
]{revtex4-2}

\usepackage{dcolumn}

\usepackage{bm}
\usepackage{cancel}
\usepackage[dvipsnames]{xcolor} 
\usepackage{amsmath}
\usepackage{soul}

\usepackage{listings}

\usepackage{hyperref}
\hypersetup{
    colorlinks=true,
    citecolor=orange,
    linkcolor=teal,
    filecolor=magenta,      
    urlcolor=teal,
    }
\usepackage[draft,inline,nomargin,index]{fixme}

\usepackage[utf8]{inputenc}
\usepackage[T1]{fontenc}

\usepackage{graphicx}
\usepackage{subcaption}
\usepackage{caption}
\usepackage{ragged2e}
\DeclareCaptionFormat{justified}{\justifying #1#2#3\par}
\newcommand\mnras{MNRAS}
\newcommand\pasj{PASJ}

\newcommand{\lcdm}{$\Lambda \text{CDM }$}

\newcommand{\Omm}{\Omega_m}

\newcommand{\post}{\mathcal{P}}

\newcommand{\prior}{\text{prior}}

\newcommand{\ellmax}{\ell_\text{max}} 
\newcommand{\ellN}{\ell_\text{N}} 

\newcommand{\ngal}[1]{n_{\text{gal},#1}} 

\newcommand{\grms}{\gamma_{\text{rms}}}

\newcommand{\kappag}{\kappa^\text{G}}

\newcommand{\sigmag}{\sigma^G}
\newcommand{\xig}{\xi^G}
\newcommand{\Cg}{C^G}

\newcommand{\gammasys}{\gamma^\text{pix}}
\newcommand{\kappasys}{\kappa^\text{pix}}
\newcommand{\gammaobs}{\gamma^\text{obs}}

\newcommand{\fskappasmooth}{{\fs\kappa}^\text{smooth}}

\newcommand{\fskappaalias}{{\fs\kappa}^\text{alias}}
\newcommand{\Calias}{C^\text{alias}}
\newcommand{\gammatrue}{\gamma^\text{true}}
\newcommand{\kappatrue}{\kappa^\text{true}}

\newcommand{\fskappatrue}{\fs\kappa^\text{true}}

\newcommand{\LN}{\mathbb{L}}
\newcommand{\KS}[1]{\mathbb{K}_{#1}}

\newcommand{\healpix}{\texttt{HEALPix}}
\newcommand{\nside}{\texttt{nside}}

\newcommand{\flask}{\texttt{FLASK}}
\newcommand{\camb}{\texttt{camb}}

\newcommand{\jax}{\texttt{jax}}
\newcommand{\numpyro}{\texttt{numpyro}}
\newcommand{\miko}{\texttt{Miko}}

\newcommand{\Mpc}{\text{Mpc}}
\newcommand{\degc}{^\circ}

\newcommand{\arcmin}{'}

\newcommand{\Sec}{\S ~}
\newcommand{\App}{App.~}
\newcommand{\Eqn}{Eq.~}
\newcommand{\Eq}{Eq.~}
\newcommand{\Eqs}{Eqs.~}
\newcommand{\Fig}{Fig.~}

\newcommand{\response}[1]{{\textcolor{black}{#1}}}

\DeclareFontFamily{U}{wncy}{}
\DeclareFontShape{U}{wncy}{m}{n}{<->wncyr10}{}
\DeclareSymbolFont{mcy}{U}{wncy}{m}{n}
\DeclareMathSymbol{\Sh}{\mathord}{mcy}{"58}

\newcommand{\sinc}{\text{sinc}}

\newcommand{\fs}[1]{\Tilde{#1}}

\newcommand{\HT}[1]{\mathcal{H}\left[#1\right]}
\newcommand{\iHT}[1]{\mathcal{H}^{-1}\left[#1\right]}

\newcommand{\mean}[1]{\langle #1 \rangle}
\newcommand{\expect}[1]{\langle #1 \rangle}

\newcommand{\bfc}{{\mathbf{c}}}
\newcommand{\bfd}{{\mathbf{d}}}

\newcommand{\bfs}{{\mathbf{s}}}

\newcommand{\bfx}{{\mathbf{x}}}

\newcommand{\bfalpha}{{\pmb{\alpha}}}
\newcommand{\bfbeta}{{\pmb{\beta}}}

\newcommand{\bftheta}{{\pmb{\theta}}}

\newcommand{\bfell}{{\pmb{\ell}}}

\newcommand{\Z}{\mathbb{Z}}

\newcommand\be{\begin{equation}}

\def\ee{\end{equation}}
\def\bea{\begin{eqnarray}}
\def\eea{\end{eqnarray}}

\begin{document}
\preprint{APS/123-QED}
\title{Accurate field-level weak lensing inference for precision cosmology}
\author{Alan Junzhe Zhou}
\email{junzhez@andrew.cmu.edu}
\author{Xiangchong Li}
\author{Scott Dodelson}
\author{Rachel Mandelbaum}
 \affiliation{Department of Physics, Carnegie Mellon University, Pittsburgh, PA 15213.\\
 McWilliams Center for Cosmology, Carnegie Mellon University, Pittsburgh, PA 15213.\\
 NSF AI Planning Institute, Carnegie Mellon University, Pittsburgh, PA 15213.
}
\date{\today}

\begin{abstract}
We present \miko{}, a catalog-to-cosmology pipeline for general flat-sky field-level inference, which provides access to cosmological information beyond the two-point statistics. In the context of weak lensing, we identify several new field-level analysis systematics (such as aliasing, Fourier mode-coupling, and density-induced shape noise), quantify their impact on cosmological constraints, and correct the biases to a percent level. Next, we find that model misspecification can lead to both absolute bias and incorrect uncertainty quantification for the inferred cosmological parameters in realistic simulations. The Gaussian map prior infers unbiased cosmological parameters, regardless of the true data distribution, but it yields overconfident uncertainties. The log-normal map prior quantifies the uncertainties accurately, although it requires careful calibration of the shift parameters for unbiased cosmological parameters. We demonstrate systematics control down to the $2\%$ level for both models, making them suitable for ongoing weak lensing surveys.
\end{abstract}
\maketitle

\section{INTRODUCTION}
\label{sec:intro}

A cosmological field-level inference (FLI) compares the field-level observables (such as galaxy density \cite{jasche_bayesian_2013,jasche_past_2015,lavaux_unmasking_2016,zhou_field-level_2023, stadler_cosmology_2023, andrews_bayesian_2022,schmidt_n-th_2021,schmidt_rigorous_2019,elsner_cosmology_2020}, cosmic shear \cite{alsing_cosmological_2017,fiedorowicz_karmma_2022-1,loureiro_almanac_2022,porqueres_field-level_2023,boruah_map-based_2022}, or cosmic microwave background (CMB) maps \cite{millea_bayesian_2020,millea_optimal_2021,zhou_field-level_2023}) directly to theoretical predictions or simulations. 
It treats the map pixels and cosmological parameters as random variables to be jointly modeled, with map-making and cosmological parameter inference being carried out simultaneously in a Bayesian manner. 

Compared to traditional two-point analyses, FLI has several advantages:
\begin{enumerate}
    \item In principle, it utilizes all the $N$-point statistics in the data, so there is minimal information loss
    \item It is built around a conceptually simple forward model which easily extends to new physics and systematics
    \item It combines the map-making and parameter inference into a single step, enabling joint inference of multiple surveys on the map level
    \item It avoids the needs the compute complex high-dimensional covariance matrices
\end{enumerate}
Previous works \cite{boruah_map-based_2023,leclercq_accuracy_2021,porqueres_field-level_2023} have shown that FLI can achieve superior cosmological constraints compared to two-point analyses. The goal of this work is to establish FLI methods with rigorously demonstrated systematics control at least at the level needed for current imaging surveys.
\be
\post(\bfc, \bfs| \bfd) \propto G(\bfd-\bfs; 0, N) \,\times \post(\bfs| \bfc) \,\times\,\prior(\bfc) \,.
\ee
Here, the set of cosmological and astrophysical parameters are denoted $\bfc$; the map pixel values (the {\it signal}, $\bfs$; the data pixel values $\bfd$; and $G(\bfx;\mu,N)$ is a Gaussian distribution of $\bfx$ with mean $\mu$ and variance $N$. We will treat the noise as uncorrelated so that the matrix $N$ is diagonal (in the case of lensing, related to the shape noise). The second probability on the right encodes two broad areas of complexity. First, even in the simplest case, where the observations are on the largest of scales so the fundamental fields are Gaussian, care needs to be taken to convert the true field to $\bfs$, the field that will be compared to the data \cite{nguyen_impacts_2021}. We call this set of issues {\it analysis systematics}, and present our treatment of them in \Sec \ref{sec:analysissys} below. Second, most observations are not sampling purely Gaussian fields, so understanding the distribution from which the true fields are drawn is paramount. We call this uncertainty {\it model misspecification} and discuss this in \Sec \ref{sec:modelmisspecs} below. This paper aims to make two main points:
\begin{enumerate}
    \item We have developed a pipeline to analyze weak lensing data in the flat sky case\footnote{The flat sky case has the advantage of requiring less computational resources and also applicable to deep surveys that cover only a fraction of the sky. Indeed, we will test our methods on realistic mocks that simulate the Hyper Suprime-Cam SSP Survey~\citep{HSC_SSP2018} year-3 shape catalog~\cite{HSC3_catalog}, and the pipeline developed here may be applicable to the Roman Space Telescope \citep{Roman2015}, current baseline wide-field imaging survey will cover about 4\% of the sky.} that includes and quantifies analysis systematics such as pixelization effects, boundary conditions, and general issues that arise when moving between real and Fourier space.
    \item In weak lensing, the choice of a Gaussian prior for $\post(\bfs| \bfc)$ leads to unbiased means on the extracted parameters but incorrect error bars. Meanwhile, using a log-normal prior for $\post(\bfs| \bfc)$ leads to correct errors bar but the means of the extracted parameters are very sensitive to the exact parametrization of the log-normal distribution. 
\end{enumerate}

 Looking forward (\Sec\ref{sec:inference}), we will constrain the tomographic power spectrum amplitudes $A_i$, which is similar to a tomographically decomposed version of $S_8$ in standard cosmological analyses. The HSC Year 3 real \cite{li_hyper_2023} and Fourier \cite{dalal_hyper_2023} space shear analyses both yield an approximately $4.1\%$ constraints on $S_8$ ($0.769^{+0.031}_{-0.034}$ and $0.776^{+0.032}_{-0.033}$ respectively). This translates to approximately an $8\%$ constraint on $A_i$. Since we typically require the systematic uncertainty to be under $1/4$ of the total error budget, we set a target of $2\%$ for the absolute bias induced by systematic effects. Given the above numbered points, we chart out a pathway for analyses that overcomes these difficulties at this target 2\% level.

This paper is organized as follows. We briefly review the basics of weak lensing in \Sec\ref{sec:wlbasics}, describe the simulated datasets in \Sec\ref{sec:sim} and then give an overview of the inference pipeline in \Sec\ref{sec:inference}.
The first set of results is presented in \Sec \ref{sec:analysissys}, which deals with analysis systematics, and the second in \Sec \ref{sec:modelmisspecs}, where model misspecification is discussed. We summarize and outlook in \Sec\ref{sec:summary}.

\noindent\rule{8cm}{0.4pt}

\section{Weak lensing basics}
\label{sec:wlbasics}
\subsection{Weak lensing maps}
As the light from distant galaxies travels toward us, it is deflected by the intervening matter overdensity perturbation $\delta_m$. Consider a line-of-sight (LOS) in the $\bftheta$ direction and a light source at comoving distance $\chi$. The light source, although emitted at the true position $\bfalpha$, is observed at $\bfbeta$ due to gravitational lensing. In the weak lensing regime, the effect of this distortion is approximately linear
\begin{equation}
    \frac{\partial \bfbeta}{\partial \bfalpha}\Bigr\rvert_{\bftheta,\chi}
    = \begin{pmatrix}
        1-\kappa(\bftheta,\chi)-\gamma_1(\bftheta,\chi) & -\gamma_2(\bftheta,\chi)                        \\
        -\gamma_2(\bftheta,\chi)                        & 1-\kappa(\bftheta,\chi)+\gamma_1(\bftheta,\chi)
    \end{pmatrix}\,,
\end{equation}
where $\kappa(\bftheta,\chi)$ and $\gamma(\bftheta,\chi) = \gamma_1(\bftheta,\chi) + i \gamma_2(\bftheta,\chi)$ are called the convergence and the shear maps \cite{bartelmann_weak_2001}.

In a tomographic weak-lensing survey, we categorize the source galaxies into redshift bins with normalized galaxy density distributions $\ngal{i}(z)$. We are interested in the line-of-sight (LOS)-averaged convergence map, which, for a spatially flat universe, is related to the matter overdensity via
\begin{align}
    \kappa_i(\bftheta) & = \int^{\chi^*}_0 d\chi W_i(\chi) \delta_m(\chi\bftheta,\chi)\,,                                                  \\
    W_i(\chi)          & = \frac{3}{2}H_0^2\Omm(1+z(\chi))\int_\chi^{\chi^*}d\chi'n_{\text{gal},i}(\chi')\frac{\chi(\chi'-\chi)}{\chi'}\,,
\end{align}
where $H_0$ is the Hubble constant, $\Omm = 8 \pi G \rho_m / (3 H_0^2)$, $\rho_m$ is the matter density, and $\chi^*$ is the comoving horizon.

\subsection{Weak lensing statistics}
The $n$-point correlation functions of the convergence field carry significant cosmological information. The two-point cross-power spectrum between redshift bins $i$ and $j$ is given by (under Limber's approximation \cite{limber_analysis_1953})
\begin{equation}
    \label{eqn:2pt}
    C_{ij}(\ell) =
    \int_0^{\chi^*} d\chi \frac{W_i(\chi)W_j(\chi)}{\chi} P_\delta(k=\frac{\ell}{\chi};\chi) \,,
\end{equation}
where $P_\delta(k;\chi)$ is the 3-dimensional matter power spectrum. In practice, we compute \Eq\ref{eqn:2pt} using \camb{} \cite{lewis_efficient_2000}. 

\begin{figure*}
    \centering
    \includegraphics[width=0.9\hsize]{ 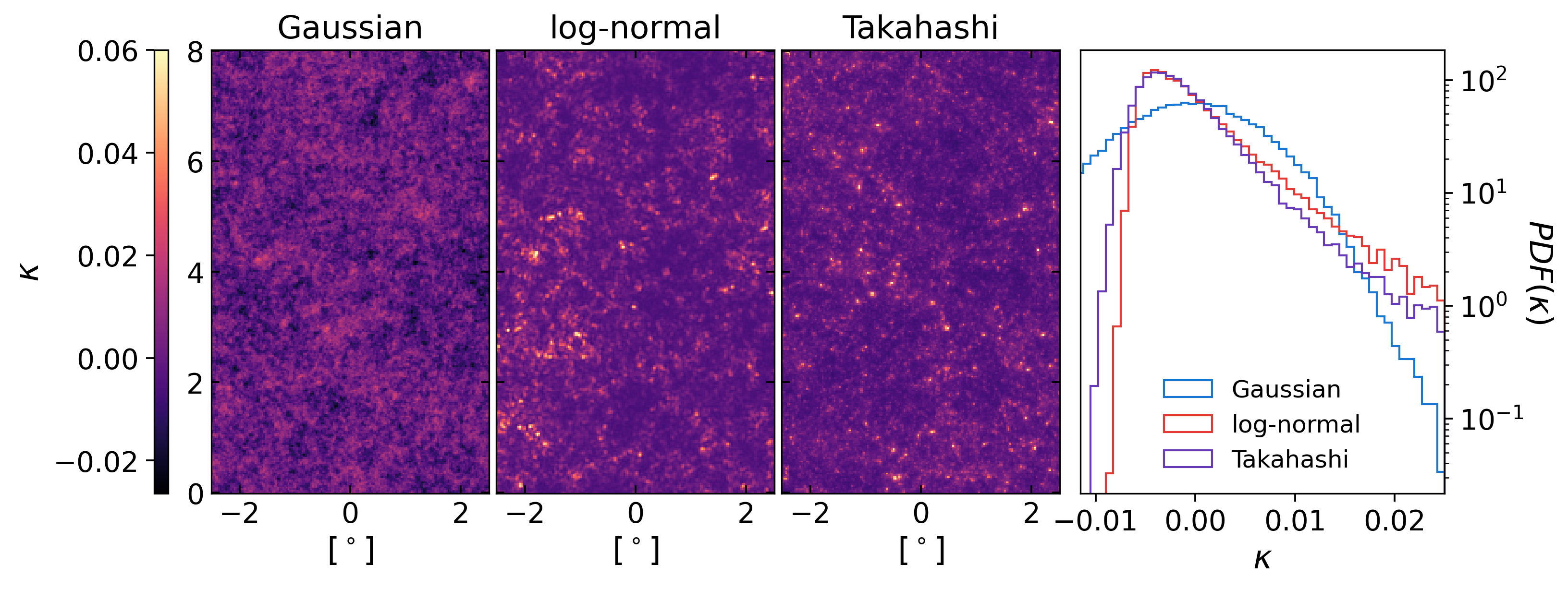}
    \caption{Examples of Gaussian, log-normal, and Takahashi fields. All three fields have the same two-point statistics (modeled after the first redshift bin \Sec\ref{sec:sim}) but a visibly different field-level appearance. Their one-point PDFs are shown in the right plot, where we see that the log-normal field has a stronger cutoff at small $\kappa$ compared to the others. We limit the maximum color map scale to $\kappa=0.06$ to emphasize the visual distinction between the fields.}
    \label{fig:field_example}
\end{figure*}

If the convergence field were completely Gaussian, then \Eqn\ref{eqn:2pt} would contain all the information. In reality, however, small-scale structure formation leads to non-Gaussianity characterized by extended voids and density peaks that contains important cosmological information \cite{liu_cosmology_2015,li_constraining_2019}. For example, \Fig\ref{fig:field_example} shows three examples of convergence fields generated with models (Gaussian, log-normal, and realistic $N$-body ray-tracing simulation). Although these fields all share the same power spectrum, they are visibly different, each with its own one-point PDF and higher-order statistics. In this work, we model the field-level statistics to extract information not captured by the standard two-point analyses.

One of the most studied models of the convergence PDF is the log-normal distribution \cite{bohm_bayesian_2017,hilbert_cosmic_2011,xavier_improving_2016,boruah_map-based_2022}. Let $\kappag$ be a zero-mean Gaussian (transfer) field. The associated zero-mean log-normal field is defined by applying a transformation across the entire Gaussian field,
\begin{equation}
    \label{eqn:lognormal_transformation}
    \kappa = \LN(\kappag,a)
    = a \left(e^{\kappag-{\sigmag}^2/2}-1 \right)\,,
\end{equation}
where $a$ is the shift parameter and $\sigmag$ is the standard deviation of the transfer map. The log-normal field is fully characterized by its variance $\sigma^2$ and shift parameter, giving it one more degree of freedom than the Gaussian field. The shift parameter $a$ is particularly interesting, as it corresponds to the minimum value of the log-normal distribution. Alternatively, $a$ can be reparametrized in terms of the skewness $s$ \cite{xavier_improving_2016} by
\begin{align}
    \label{eqn:shift_skewness}
    a    & = \frac{\sigma}{s}\left(1+y(s)+y^{-1}(s)\right) \,,   \\
    y(s) & = \left(\frac{2+s^2+s\sqrt{4+s^2}}{2}\right)^{1/3}\,, \\
    s    & = \expect{(\kappa/\sigma)^3} \,,
\end{align}
which we find to be a more robust measurement on real simulations. The assumption about $a$ is very important and we will discuss it in \Sec\ref{sec:lognormal_result}.

Finally, we will need to relate the two-point correlation function of the transfer field to that of the log-normal field. This is given by
\begin{equation}
    \xig_{ij}(\theta) = \log\left(\frac{\xi_{ij}(\theta)}{a_i a_j}+1\right)\,.
\end{equation}
The two-point correlation function is related to the power spectrum by the classic Hankel transform
\begin{align}
    \xi_{ij}(\theta)
     & = \iHT{C_{ij}(\ell)}(\theta)                                     \\
     & = 2\pi \int d\theta ~ \theta ~ C_{ij}(\ell) ~ J_0(\ell\theta)\,,
\end{align}
where $J_0$ is the Bessel function of the first kind.

\subsection{Weak lensing observables}
A weak lensing survey measures the shapes of galaxies to infer the convergence and/or the shear \cite{mandelbaum_third_2014}. For a galaxy with observed complex shear $\gamma_o$,
\begin{equation}
    \label{eqn:obs_shear}
    \gamma_{o} = \gamma + \gamma_{I} + \gamma_{n}\,.
\end{equation}
The first term on the right hand side is the lensing shear, $\gamma_I$ is the
coherent distortion of galaxy shape due to intrinsic alignment (IA) \cite{troxel_intrinsic_2015,tsaprazi_field-level_2022}, and $\gamma_n$ is the galaxy shape noise caused by random intrinsic galaxy shape and image noise. In this study, we leave $\gamma_{I} = 0$ and model $\gamma_{n}$ as a zero-mean Gaussian random variable with standard deviation $\grms$. Additionally, higher-order differences between reduced shear and shear \citep{DESY3_highOrder_Krause2021} are not considered in our analysis.

\section{SIMULATIONS AND DATASETS}
\label{sec:sim}
We model our analysis after the HSC Year 3 survey specifications \cite{HSC3_catalog, rau_weak_2023}. We use three sets of independent simulations to test our analysis pipeline across the entire galaxy catalog to cosmological constraints process. In this section, we discuss the process of converting galaxy catalog to the observed pixelized shear maps $\gammaobs_i$. We will show in \Sec\ref{sec:analysissys} that this procedure introduces significant statistical artefacts in $\gammaobs_i$ that must be reflected in our forward model for unbiased cosmological constraints.

\begin{figure}[!htbp]
    \centering
    \includegraphics[width=1\hsize]{ 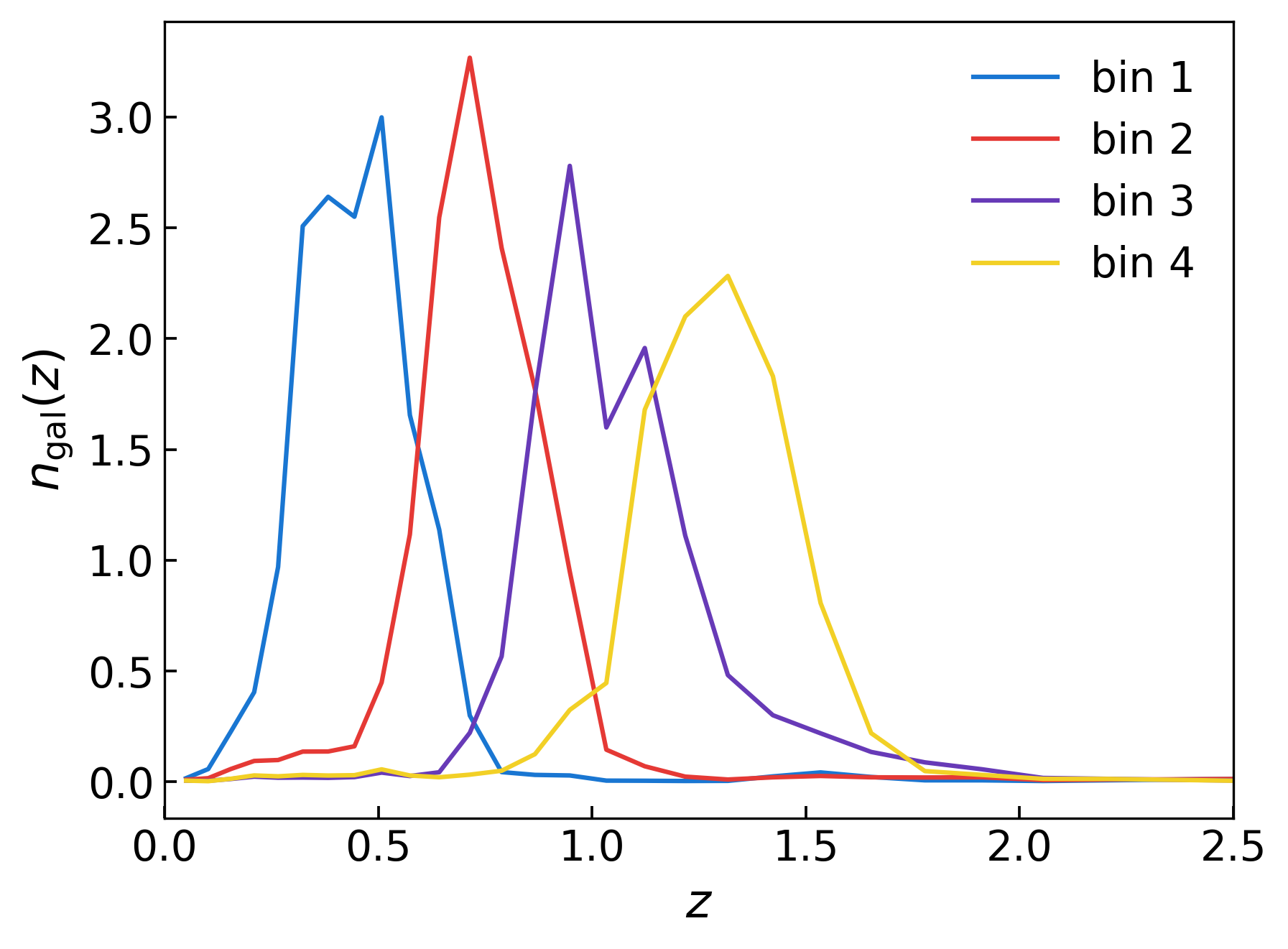}
    \caption{The normalized PDF of the tomographic redshift distributions of the HSC Year 3 galaxy shape catalog \cite{rau_weak_2023}.
    }
    \label{fig:nz}
\end{figure}

The first simulation comes from \citet{takahashi_full-sky_2017}, where the authors perform multiple-lens plane ray-tracing on high-resolution $N$-body simulation. This simulation uses the WMAP 9 years result \cite{hinshaw_nine-year_2013} as its fiducial cosmology, which we also adopt through out our work. The Takahashi simulation has been extensively tested and used in HSC to derive covariance of 2-point summary statics \cite{HSC1_mock_Shirasaki2019}. It provides independent full sky realizations of $\kappa(\bftheta,\chi)$ and $\gamma(\bftheta,\chi)$ at the $38$ comoving radial shells. These maps are provided in \healpix{}, format \cite{gorski_healpix_2005,zonca_healpy_2019} with a resolution of $\nside=8192$\,. We use the radial shells to construct the 4 tomographic $\kappa_i(\bftheta)$ and $\gamma_i(\bftheta)$ maps using the HSC Year 3 redshift distribution (\Fig\ref{fig:nz}). Next, we locate rectangular patches of side lengths $(L_x, L_y) = (8\degc, 5 \degc)$ and the patches are separated by at least $8\degc$ to avoid spatial correlations. We generate galaxies with tomographic effective number density $\rho_i$ with true $\kappa$ and $\gamma$ values according to their positions. 

We set $\rho_i$ to be $36.1$, $52.0$, $39.6$, and $20.6 ~ \mathrm{arcmin}^{-2}$ for the four redshift bins, corresponding to $10$ times the HSC Year 3 W12H field specification (explained below). Then, we add Gaussian shape noise to individual galaxies according to \Eq\ref{eqn:obs_shear} with $\grms=0.26$ \cite{LSSTRequirement2018}. We construct the pixelized maps $\kappatrue_i$ and $\gammatrue_i$ using the noiseless catalog, and construct $\gammaobs_i$ using the noisy catalog. In both cases, the pixels have size $\Delta = 3 \arcmin$.

The choice of $\rho_i$ merits a few comments. For a pixel $p$, the effective shape noise has variance
\begin{equation}
    \label{eqn:shape_noise}
    N_i(p) = \frac{\grms^2}{\rho_i(p)\ \Delta^2}\,.
\end{equation}
We are mostly interested in the bias of modeling choices and the robustness of our posterior estimation in the presence of noise and cosmic variance. Thus, given a finite computational resource, we want to reduce the statistical uncertainty and perform as many independent HSC-like experiments as possible. We achieve this goal by setting $\rho_i$ to $10$ times the fiducial HSC number density \cite{HSC3_catalog} to reduce the effective shape noise.

Due to simulation artefacts \cite{takahashi_full-sky_2017}, the Takahashi convergence maps does not recover the theoretical power spectrum (\Eq\ref{eqn:2pt}) with the precision required for this work. Therefore, we need calibrate the theory when performing inference on the Takahashi mocks. The exact procedure is detailed in \App\ref{app:cl_calibration}. The calibrated theoretical power spectrum agrees with the simulation within $1\%$ for $\ell<24000$ for all redshift bins.

The second and third sets of simulations are conceptually similar. Using the same WMAP9 cosmology, we make full sky log-normal and Gaussian realizations of convergence and shear at resolution of $\nside=8192$, and then construct shear catalogs and flat sky maps as before. We generate the log-normal maps using \flask{} \cite{xavier_improving_2016}. For these maps, we observe a larger-than-$1\%$ surplus of power above $\ell=5000$. Therefore, we only use $\ell \leq 5000$ for both the log-normal maps and the log-normal theory power in the analyses.

For all three simulations, we use $4$ independent full-sky simulations to generate $80$ nonoverlapping flat-sky patches. 

\section{INFERENCE FRAMEWORK}
\label{sec:inference}
\begin{figure}[!htbp]
    \centering
    \includegraphics[width=1\hsize]{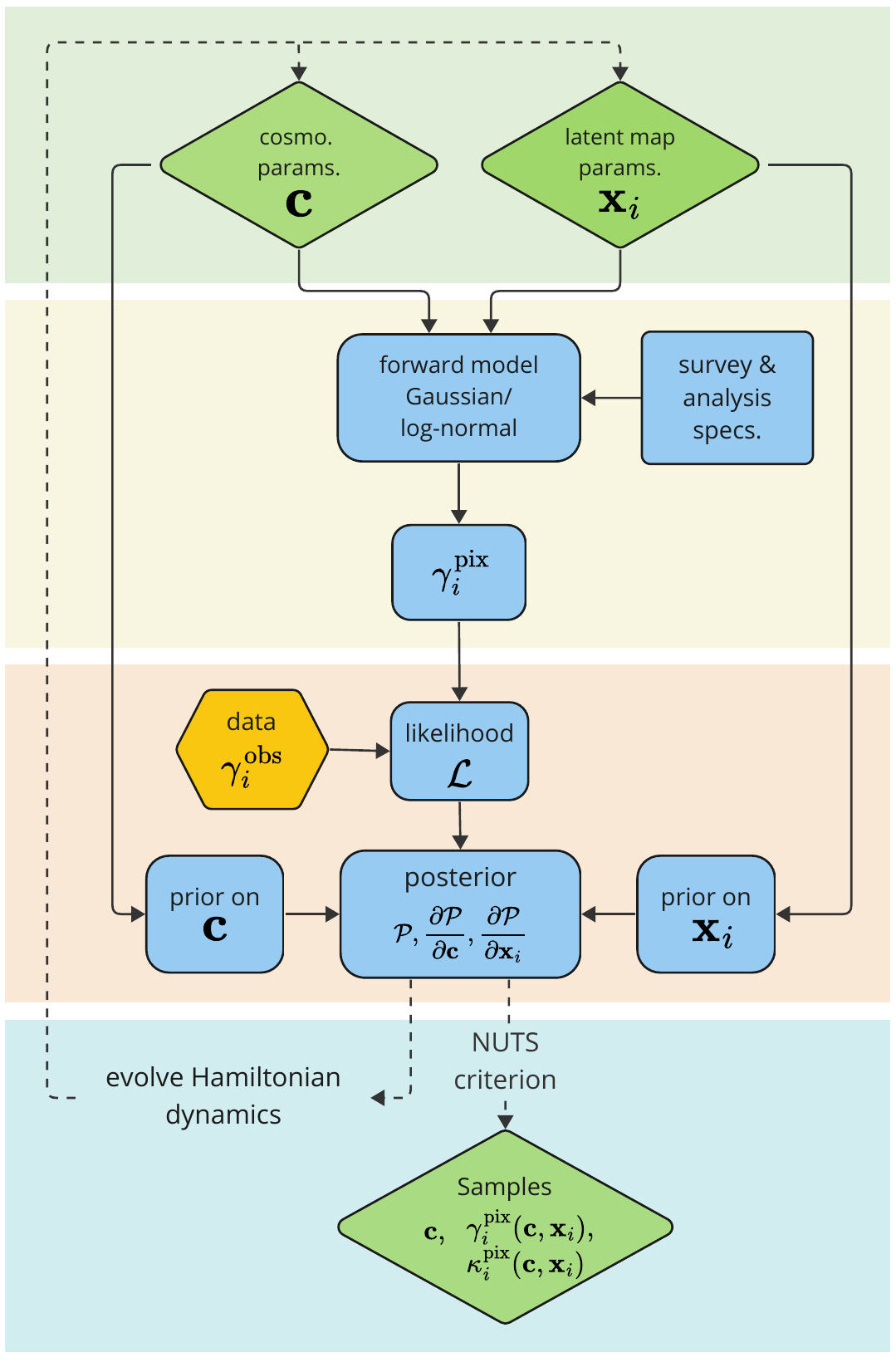}
    \caption{This flow chart describes \miko{}'s weak lensing field level forward model and inference pipeline (\Sec\ref{sec:inference}). The pipeline has four steps (from top to bottom): generating latent parameters, forward modeling the noiseless fields, computing the likelihood given the data, and sampling. In the chart, diamonds represent sampled parameters, squares the model-relevant functions, and the hexagon the observed data vector. 
    }
    \label{fig:flowchart}
\end{figure}

Our inference framework \miko, graphically illustrated in \Fig\ref{fig:flowchart}, is a hierarchical Bayesian network that forward models from the cosmological parameter to the observed noisy fields. We focus on the lensing convergence and shear fields in this work, but the framework is set up to include galaxy density and CMB fields as well \cite{zhou_field-level_2023}. The weak lensing pipeline includes two models; the first generates the observable fields using a Gaussian map prior, and the second uses a log-normal prior.

\miko~ has four main components. Starting from the top of the flow chart, we first draw cosmological parameters $\bfc$ and the latent tomographic map parameters $\bfx_i$ from their prior distributions. For the Gaussian model, we set the free cosmological parameters to be the amplitudes of the tomographic power spectra,
\begin{equation}
    \label{eqn:2pt_amp}
    C_{ij} \longrightarrow A_i A_j C_{ij}\,.
\end{equation}
where $C_{ij}$ is the power spectrum at the fiducial cosmology calibrated to the simulations\footnote{In real observations, we can simply set $C_{ij}$ to the prediction of the fiducial cosmology.} (\Sec\ref{sec:sim}). Inference of $A_i$ is a powerful test of the \lcdm model. Consider the redshift-modulated growth function $A(z)D(z)$. In this context, $A_i$ can be thought of as $A(z)$ integrated over the tomographic window function $W_i(z)$. Thus, a detection of $A_i \neq 1$ implies that the growth history near the redshift interval of bin $i$ deviates from the \lcdm model. Through out this study, we use a broad and uniform prior on $A_i$ over the interval $\left[0.5, 1.5\right]$. For the log-normal model, the free parameters are both $A_i$ and the shift parameters $a_i$. The prior on $a_i$ is more complicated and we will delay its discussion until \Sec\ref{sec:modelmisspecs}. The latent map parameters $\bfx_i$ represent the field-level degrees of freedom and are drawn from the unit normal distribution.

Next, we use the cosmological parameters and survey and analysis specifications to define a forward model, which transforms $\bfx_i$ to noiseless and pixelized tomographic convergence and shear maps $\kappasys_i$ and $\gammasys_i$ on one (or more) flat sky patch(es). It is important to note that the process of generating $\kappasys_i$ and $\gammasys_i$ includes not only the cosmological model but also the survey and analysis-specific systematics model. The process of going from $\bfx_i$ to $\gammasys$ is explained in detail in \Sec\ref{sec:analysissys}). We then compare the noiseless $\gammasys_i$ to the data $\gammaobs_i$ and build the Bayesian posterior function given by
\begin{align}
    \nonumber
     & \log \post\left(\{\gammasys_i(\bfc,\bfx_i)\}|\{\gammaobs_i\}\right) \\
     \label{eq:likelihood}
     & =
    - \sum_i \frac{\left(\gammaobs_i-\gammasys_i(\bfc,\bfx_i)\right)^2}{2N}
    + \log \prior \left(\{\gammasys_i(\bfc,\bfx_i)\}\right)\,,
\end{align}
where $N$ is the shape noise variance map given by \Eq\ref{eqn:shape_noise}. We implement all computations using differentiable programming in \jax.

In the last step, we connect this differentiable posterior function to the \numpyro{}\cite{phan_composable_2019} implementation of the Hamiltonian Monte Carlo No-U-Turn sampler (HMC NUTS) \cite{neal_mcmc_2011,bingham_pyro_2018,JMLR:v15:hoffman14a} to efficiently sample from the high-dimensional joint posterior space of both the cosmological and the map parameters. The result of the inference is comprised of the posterior samples of the cosmological parameters $\bfc$ and the maps $\kappasys$ and $\gammasys$. We could also obtain samples of the continuous fields (without the pixelization effect) if desired. 

We perform consistency checks on both models. This is done by first using the model to generate noisy shear maps and using the same model to infer $A_i$ from it. We use $10$ chains for each experiment for all the main analyses in this paper. Each chain is initialized at the maximum a posterior estimate calculated with a stochastic variational inference procedure parametrized with the $\delta$-distributions. We then run approximately $500$ warm-up NUTS steps followed by $100$ sample collection NUTS steps. We use a target probability of $0.7$ and a maximum tree depth of $9$ for our NUTS sampler. We also check the chain convergence for the posterior samples.

\section{Forward modeling and analysis systematics}
\label{sec:analysissys}
The first main result of this study is to present a catalog-to-cosmology pipeline and demonstrate that we can control the systematic uncertainty on $A_i$ to be within $2\%$ (a target motivated by our arguments in \S\ref{sec:intro}). We begin this section by laying out the algorithm that generates the maps. We then identify and catalog the systematics that significantly bias the cosmological results and provide the appropriate remedies. This section is a detailed view of the ``forward model'' box in \Fig\ref{fig:flowchart}.

\subsection{Map making}
\label{sec:map_making}
Let us first consider generating a set of correlated tomographic convergence maps with Gaussian priors. To do this, we first generate independent standard normal variables (which we shall call the latent map parameters) $\fs\bfx_i(\bfell)$ on the Fourier grid. Next, we transform these latent parameters to the Fourier space convergence maps $\fs\kappa_i(\bfell)$ by (we will denote all Fourier space quantities using $\fs{\cdot}~$)
\begin{equation}
    \label{eq:chol}
    \fs\kappa_i(\bfell) = L_{ij}(\ell)\fs\bfx_j(\bfell)\,,
\end{equation}
where $L_{ij}(\ell)$ is the Cholesky decomposition of $A_i A_j C_{ij}(\ell)$. In the trivial example with one tomographic bin and the amplitude fixed to its fiducial value ($A=1$), this would simply mean multiplying $\bfx(\bfell)$ by $\sqrt{C(\ell)}$, the RMS of the kappa map.

Now let us discuss the log-normal forward model. For the log-normal maps, we first need to compute the power spectrum of the Gaussian transfer map, this is given by
\begin{equation}
    \label{eq:lognormal_cl}
    \Cg_{ij} = \HT{\log\left(\frac{\iHT{A_iA_jC_{ij}}}{a_i a_j}+1\right)}
\end{equation}
where $\HT{\cdot}$ is the Hankel transform.

We then generate a Gaussian transfer map just as we did in the Gaussian model, Fourier transform it into the real space, and apply the log-normal transformation in \Eqn\ref{eqn:lognormal_transformation} to obtain the log-normal map.

\begin{figure}[htbp]
    \centering
    \includegraphics[width=1\hsize]{ 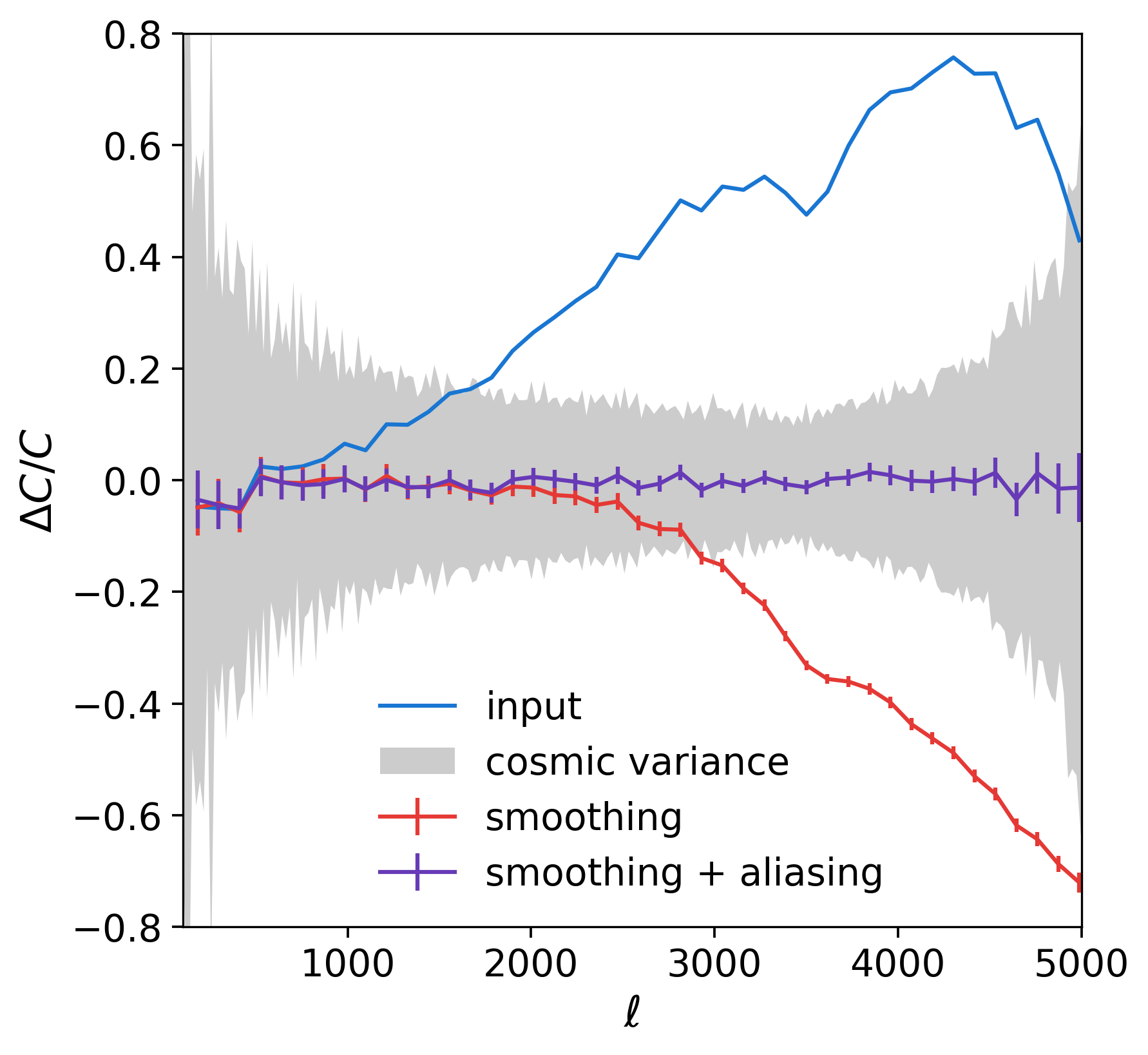}
    \caption{
    The impact of smoothing and aliasing on $C$ is presented in terms of the fractional difference relative to $\langle C(\kappatrue)\rangle$, with the gray region representing the $1\sigma$ spread of $C(\kappatrue)$. The full-sky power spectrum is shown in blue. $\mean{C(\kappasys)}$, generated by a forward model that accounts for smoothing but not aliasing, is shown in red. When both effects are included in the forward model (with aliasing parameter $r=1.6$), $\mean{C(\kappasys)}$ is depicted in purple. Error bars indicate the standard error of the mean. Neglecting either the smoothing or aliasing effects, or both, in the forward model results in an order-of-unity bias on $C$.
     }
    \label{fig:aliasing_cl}
\end{figure}

For both the Gaussian and the log-normal model, once we obtain the $\kappa_i$, we compute the shear maps using the Kaiser-Squire relationship \cite{kaiser_mapping_1993}, which is most easily expressed in the Fourier space by
\begin{align}
    \label{eqn:ks1}
    \fs\gamma_{1}(\bfell) & = \KS{1}(\fs\kappa) = \frac{\ell_x^2-\ell_y^2}{|\bfell|^2} \fs\kappa(\bfell) \\
    \label{eqn:ks2}
    \fs\gamma_{2}(\bfell) & = \KS{2}(\fs\kappa) = \frac{2\ell_x\ell_y}{|\bfell|^2} \fs\kappa(\bfell)
\end{align}

\subsection{Smoothing and aliasing}
\label{sec:smoothalias}

The process of pixelizing galaxy catalogs into shear maps (\Sec\ref{sec:sim}) introduces field-level statistical artifacts - smoothing and aliasing - in $\gammaobs$. Both effects introduce corrections to $C$ that are on the same order as $C$ itself on small scales. For example, \Fig\ref{fig:aliasing_cl} shows that the full-sky theory $C$ (blue) and the distribution of $\kappatrue$ for Gaussian mocks (gray) differ by as much as $80\%$. An unbiased forward model should generate $\kappasys$ such that $\mean{C(\kappasys)} = \mean{C(\kappatrue)}$. Neglecting the pixelization effect significantly biases the cosmological results. To the best of our knowledge, this is the first precise quantification and correction of the aliasing effect in the context of field-level cosmology.

\begin{figure*}
    \centering
    \includegraphics[width=1\hsize]{ 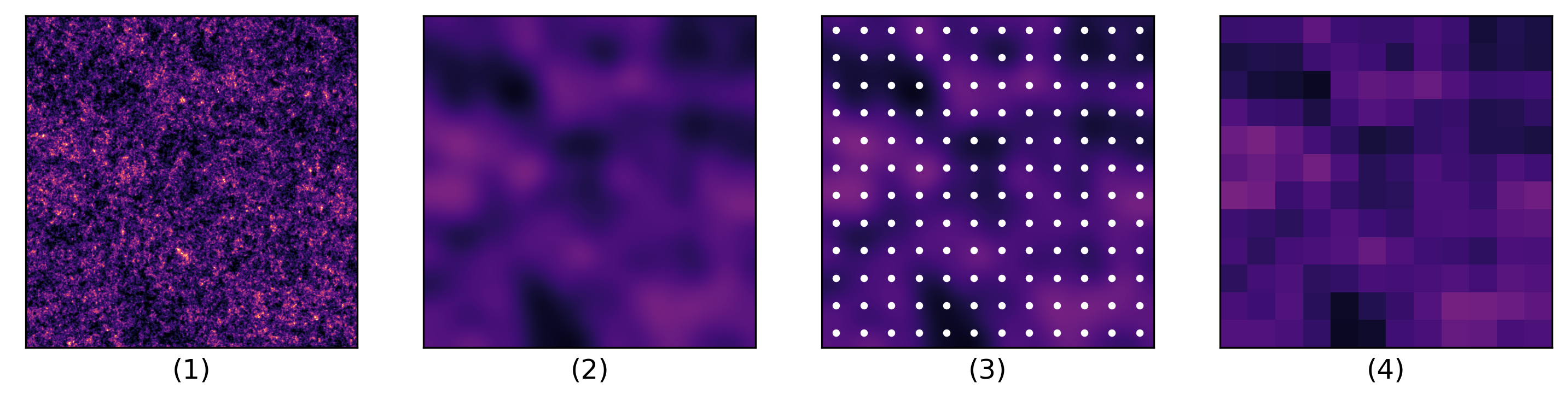}
    \caption{Illustration of the mathematical operations implied by the map-making process (\Sec\ref{sec:sim}). Plot (1) shows a continuous field. During map-making, the field is first smoothed with a pixel kernel (2). The smoothed field is then sampled at the pixel centers, indicated by white dots in (3). The values at these pixel centers constitute the pixelized map seen in (4). Both the smoothing and sampling processes introduce significant statistical artifacts into the pixelized maps, which must be accounted for in the forward model.}
    \label{fig:pixelization}
\end{figure*}

Let us start by drawing a large number of galaxies from a continuous and noiseless convergence field $\kappa$. Mathematically, the process of map-making (\Sec\ref{sec:sim}) is equivalent to first smooth $\kappa$ with a pixel kernel and then sampling it at the center of each pixel (\Fig\ref{fig:pixelization}).
Both effects are most easily described in the Fourier space. Smoothing is described by
\begin{align}
    \label{eqn:smoothing}
    \fskappasmooth
     & =
    \sinc\left(\frac{\ell_x \Delta }{2\pi}\right)
    \sinc\left(\frac{\ell_y \Delta }{2\pi}\right) ~ \fs \kappa \,,
\end{align}
Meanwhile, the innocent-looking sampling procedure not only discretizes the Fourier space but also aliases $\kappa$. The net effect increases the small-scale (near the Nyquist frequency of the observed map, $\ellN = \pi / \Delta$) power of the pixelized map by an order of unity and breaks the statistical isotropy of $\kappa$. More precisely, pixelizing the continuous field $\kappa$ means \cite{kirchner_aliasing_2005, sefusatti_accurate_2016} 
\begin{widetext}
    \begin{align}
        \label{eqn:aliasing_map}
        \fskappaalias(\bfell)
         & = (\fs\Sh * \fs\kappa)(\bfell)
        + \underbrace{\sum_{(k_1,k_2) \neq (0,0)} (\fs\Sh * \fs\kappa)(\ell_x-k_1\ellN,\ell_y-k_2\ellN)}_{\text{aliases}}\,, \\
        \label{eqn:aliasing_ps}
        \Calias(\bfell)
         & = C(\bfell)
        + \underbrace{\sum_{(k_1,k_2) \neq (0,0)}
            C_{ij}(\ell_x-k_1\ellN,\ell_y-k_2\ellN)}_{\text{aliases}}\,.
    \end{align}
\end{widetext}
We define $*$ as the (Fourier space) convolution and the Kronecker comb function as
\begin{equation}
    \Sh(\bftheta) = \sum_{n_x,n_y \in \Z}
    \delta_\text{K}\left(\theta_x - n_x \Delta \right)
    \delta_\text{K}\left(\theta_y - n_y \Delta \right)\,,
\end{equation}
where $\delta_\text{K}(x) = 1$ when $x = 0$ and $0$ otherwise.

In short, \Eq\ref{eqn:aliasing_map} tells us that each mode of $\fskappaalias(\bfell)$ is a superposition of $\fs\kappa(\bfell)$ and its ``aliases'' at higher frequencies. \Eq\ref{eqn:aliasing_ps} tells us that the power spectrum of the pixelized map acquires a two-dimensional $\bfell$ dependence and is no longer isotropic. Furthermore, $\Calias(\bfell)$ is also the superposition of $C(\bfell)$ and its aliases. One immediate consequence is that, around the Nyquist frequency, $\Calias$ is at least $100\%$ higher than $C$. Although \Eqn\ref{eqn:aliasing_map} suggests that accounting for infinitely many aliases at increasingly larger $\ell$'s is necessary to fully capture aliasing, in practice, aliases above $\sqrt{2} r \ellN$ with $r=1.6$ are sufficiently suppressed by the smoothing kernel. 

We implement \Eqs\ref{eqn:smoothing} and \ref{eqn:aliasing_map} in \miko{} to model the pixelization effect. The results are demonstrated in \Fig\ref{fig:aliasing_cl} for the Gaussian model. If we only include smoothing (red) in the forward model, the $\mean{C(\kappasys)}$ is two times lower near $\ellN$, as expected. When both smoothing and aliasing are included (purple), the model generates $\kappasys$'s that align with $\kappatrue$ on the two-point level across all scales. This behavior is consistently observed across all the cross-power spectra and is also confirmed for the log-normal model.

\begin{figure}
    \centering
    \includegraphics[width=1\hsize]{ 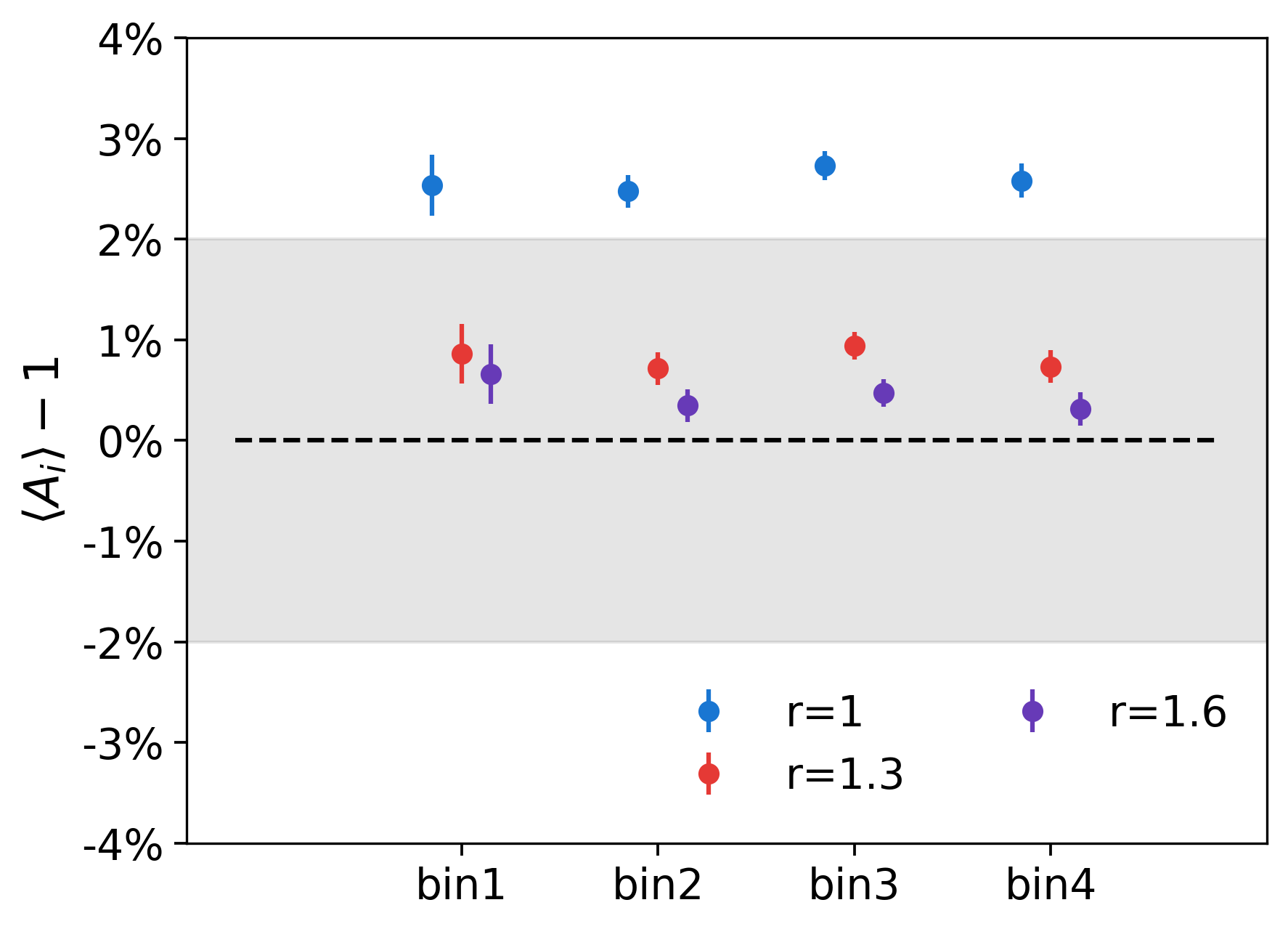}
    \caption{The absolute bias of $A_i$ measured by applying the Gaussian model to Gaussian mocks. Each color corresponds to a Gaussian model with a different aliasing parameter $r$. Each data point represents the average inferred $A_i$ over 80 independent HSC-like experiments, with the error bars denoting the standard error of the mean. The experiment is set with an $80\%$ margin (see \Sec\ref{sec:bc} and \Fig\ref{fig:A_margin_gauss}). The gray region indicates the $2\%$ systematics control target. }
    \label{fig:A_alias_gauss}
\end{figure}

We now test the bias introduced by the pixelization effect on the cosmological parameters $A_i$. We setup three different Gaussian models. Each model includes the smoothing correction but differs in the aliasing parameter $r$, set at $1$ (no aliasing correction), $1.3$, and $1.6$ respectively. We run each model on $80$ HSC-like experiments, using Gaussian mocks. The results are presented in \Fig\ref{fig:A_alias_gauss}.
Neglecting aliasing in the forward model biases $A_i$ higher. The direction of the bias is expected since aliasing amplifies the small-scale power in $\gammaobs$ which can only be compensated by increasing $A_i$. Neglecting aliasing leads to a $2.5\%$ absolute bias on the cosmological parameters exceeding our error budget. However, setting $r=1.6$ is generally sufficient to model the aliasing effect. We will use $r\ge1.6$ for all the following analyses.

\subsection{Boundary conditions}
\label{sec:bc}
\begin{figure}
    \centering
    \includegraphics[width=1\hsize]{ 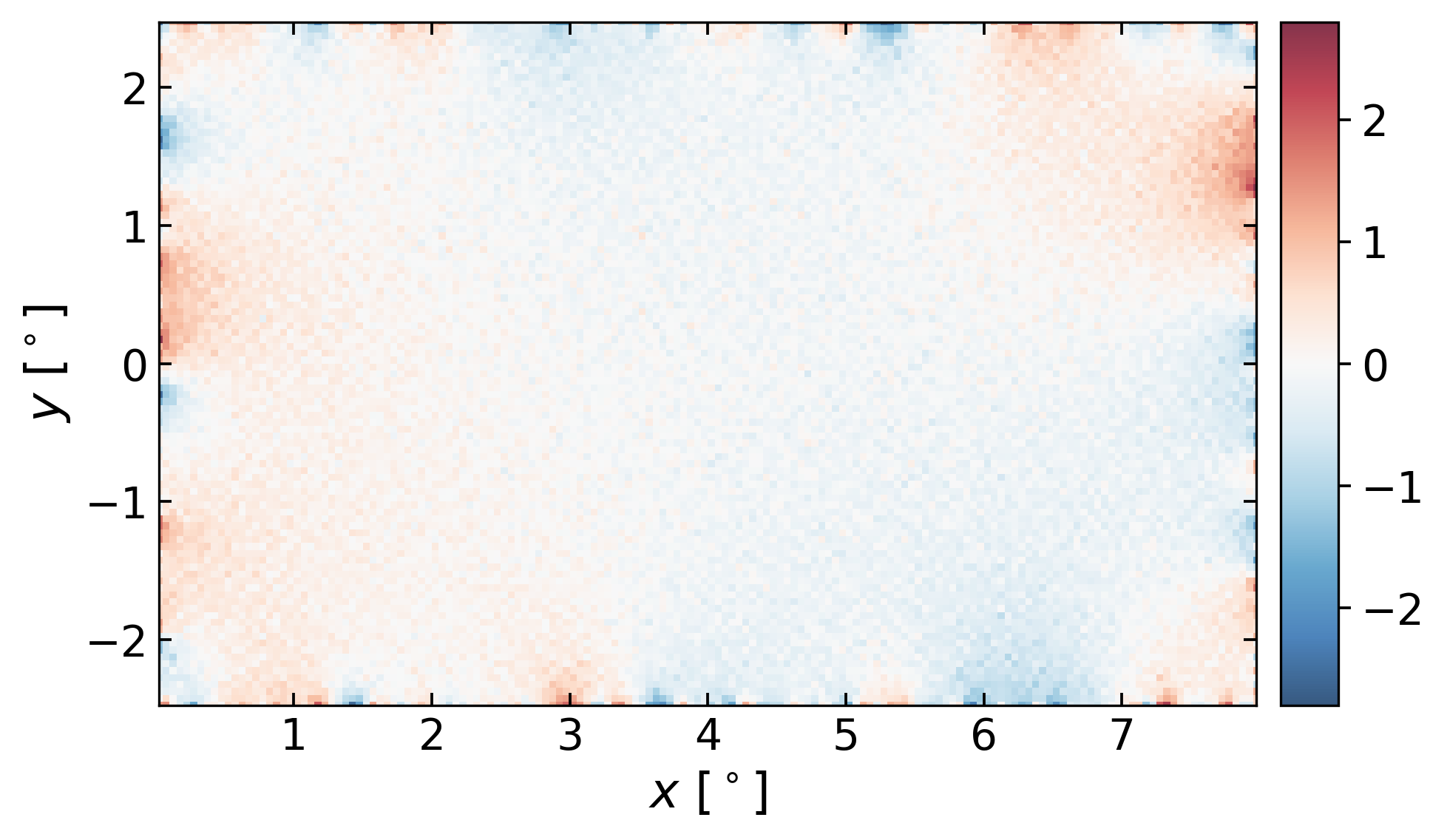}
    \caption{The Kaiser-Squire reconstruction error of a (Gaussian) shear field assuming periodic boundary condition. We define the reconstruction error as $\epsilon = (\KS{1}\kappa - \gamma_1)/\Delta_\gamma$, where $\KS{1}$ is the Kaiser-Squire operator that transforms $\kappa$ to $\gamma_1$ assuming a nonperiodic boundary at spatial infinity. $\Delta_\gamma$ is the standard deviation of the shear field. We observe the most error along the boundaries.     }
    \label{fig:kaisersquire}
\end{figure}

\begin{figure}
    \centering
    \includegraphics[width=1\hsize]{ 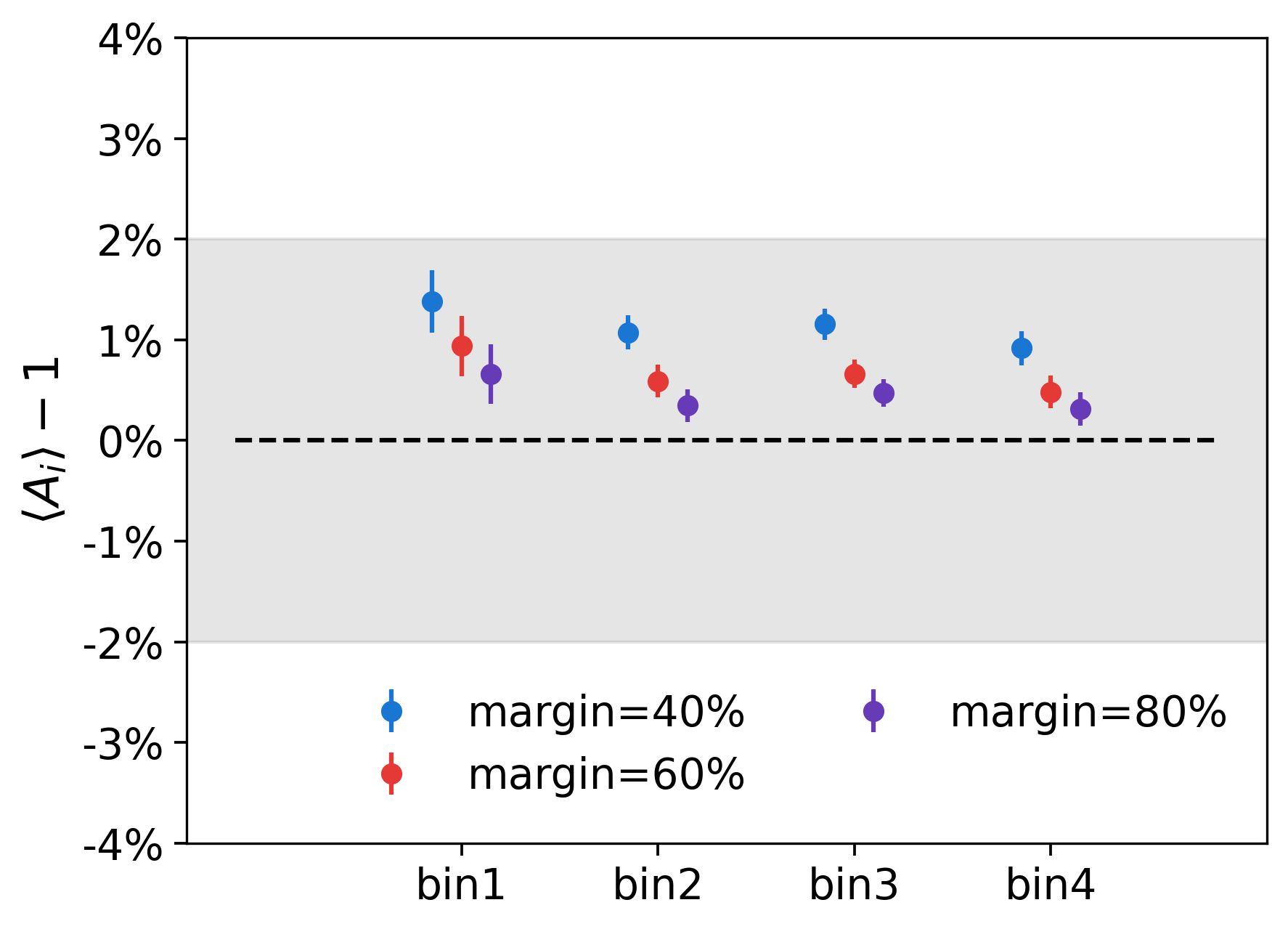}
    \caption{
    The absolute bias of $A_i$ measured by applying the Gaussian model to Gaussian mocks. Each color corresponds to a Gaussian model with a different margin size (as a fraction of the length of the short side of the patch). Each data point represents the average inferred $A_i$ over 80 independent HSC-like experiments, with the error bars denoting the standard error of the mean. The experiment is set with $r=1.6$ (see \Sec\ref{sec:smoothalias} and \Fig\ref{fig:A_alias_gauss}). The purple points corresponds to the same purple points in \Fig\ref{fig:A_alias_gauss}. The gray region indicates the $2\%$ systematics control target.}

    \label{fig:A_margin_gauss}
\end{figure}

We use the Kaiser-Squire relationship \Eqs\ref{eqn:ks1} and \ref{eqn:ks2} to transform $\kappa$ into $\gamma$ in the forward model. The Kaiser-Squire relationship is exact on an infinite plane. However, our forward model is defined on a finite grid which has a periodic boundary condition inherited from the fast Fourier transform. This mismatch introduces field-level errors around the edges of the maps as shown in \Fig\ref{fig:kaisersquire}.
We can alleviate this bias by generating maps larger than the survey footprint and defining the likelihood on the inner regions. More specifically, we test how different amounts of margins, measured by the fractional length of the short side of the map, affect the final cosmological parameter constraints. The result is shown in \Fig\ref{fig:A_margin_gauss}. We find that the boundary condition mismatch drives $A_i$ higher. Furthermore, a $80\%$ margin is sufficient to reduce the bias below $0.5\%$. We will use this choice of margin for all the following analyses.

\subsection{Fourier mode coupling}
\label{sec:modecoup}
As pointed out by \citet{xavier_improving_2016} and \citet{tessore_glass_2023}, the log-normal model transformation is not local in Fourier space and strongly couples the small-scale modes. A finite Fourier grid implies a strict band-limit and therefore changes the expected behavior of the transformation. To faithfully simulate a log-normal map up to $\ellmax$, we need to generate a Gaussian transfer map up to $\ell^G > \ellmax$. For example, $\ell^G > 4\ellmax$ is required for percent-level accuracy when simulating full-sky log-normal maps using \flask{} (as is done for our log-normal mocks \Sec\ref{sec:sim}). The iterative correction method proposed by \citet{tessore_glass_2023} can, in theory, produce logarithmic normal maps with accurate power spectra with $\ell^G \approx \ellmax$. We, however, did not achieve the same success when applying the same idea to the flat sky case, potentially due to the boundary condition and the nonisotropy of the Fourier grid. 

\begin{figure}
    \centering
    \includegraphics[width=1\hsize]{ 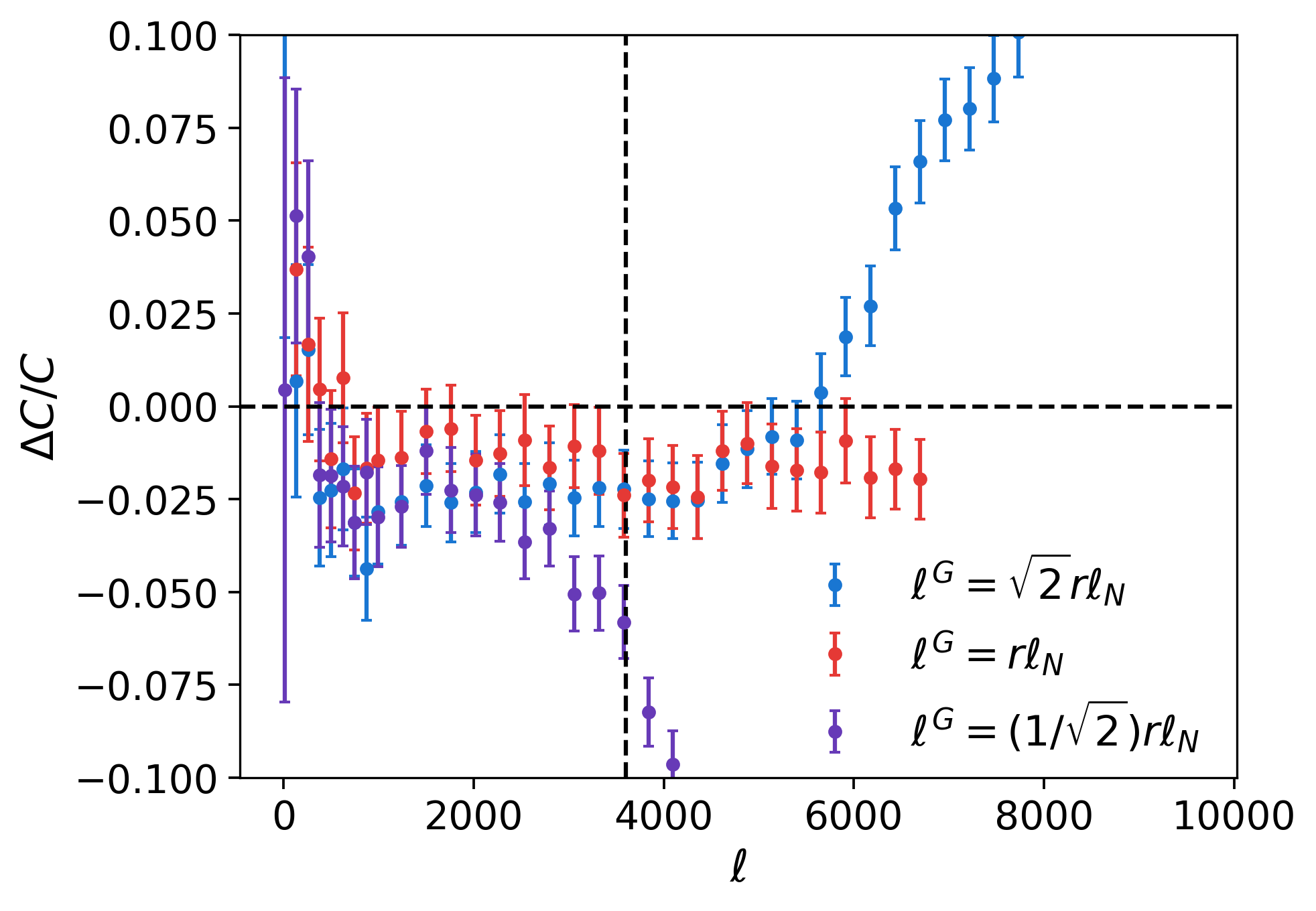}
    \caption{
    The impact of the Gaussian transfer map scale-cut, $\ell^G$, on the recovered $C$ is presented in terms of the fractional difference relative to the input $C$. The error bars represent the standard error of the mean. The vertical line denotes the Nyquist frequency, $\ellN$, of $\gammaobs$. Setting $\ell^G = r\ellN$ (red) recovers the input $C$ with the highest accuracy. The observed residual additive bias is related to the limited sky coverage of the patch.}
    \label{fig:gaussian_bandlimit}
\end{figure}

Nevertheless, we found an empirical solution to reduce this error. In \Fig\ref{fig:gaussian_bandlimit}, we plot the scale-dependent error of the log-normal $C(\kappa)$ relative to the input $C$ as a function of $\ell^G$. If we set $\ell^G$ to the maximum aliasing scale $\sqrt{2}r\ellN$ (blue), $C(\kappa)$ starts to deviate from the input beyond $\ellN$ (vertical line). To obtain the correct $C(\kappa)$ beyond $\ellN$, we find it useful to force $\kappag$ to be isotropic by removing powers beyond $r \ellN$ (red). If we impose an even stricter scale-cut, the log-normal maps exhibit lower band power below $\ellN$. Regardless of $\ell^G$, the log-normal maps still have a deficit of large-scale power. This appears to be a consequence of the small survey footprint or the periodic boundary effect, and this error is reduced when generating larger maps. For the log-normal model, since the Fourier mode coupling issues that we have discussed so far interact together, it is difficult to quantify each error separately. However, looking ahead, we will show that using $\ell^G = r\ellN$ with $r=1.9$ enables us to control the systematic bias for the log-normal model to $2\%$.

\subsection{Number density-induced shape noise}
\begin{figure}
    \centering
    \includegraphics[width=1\hsize]{ 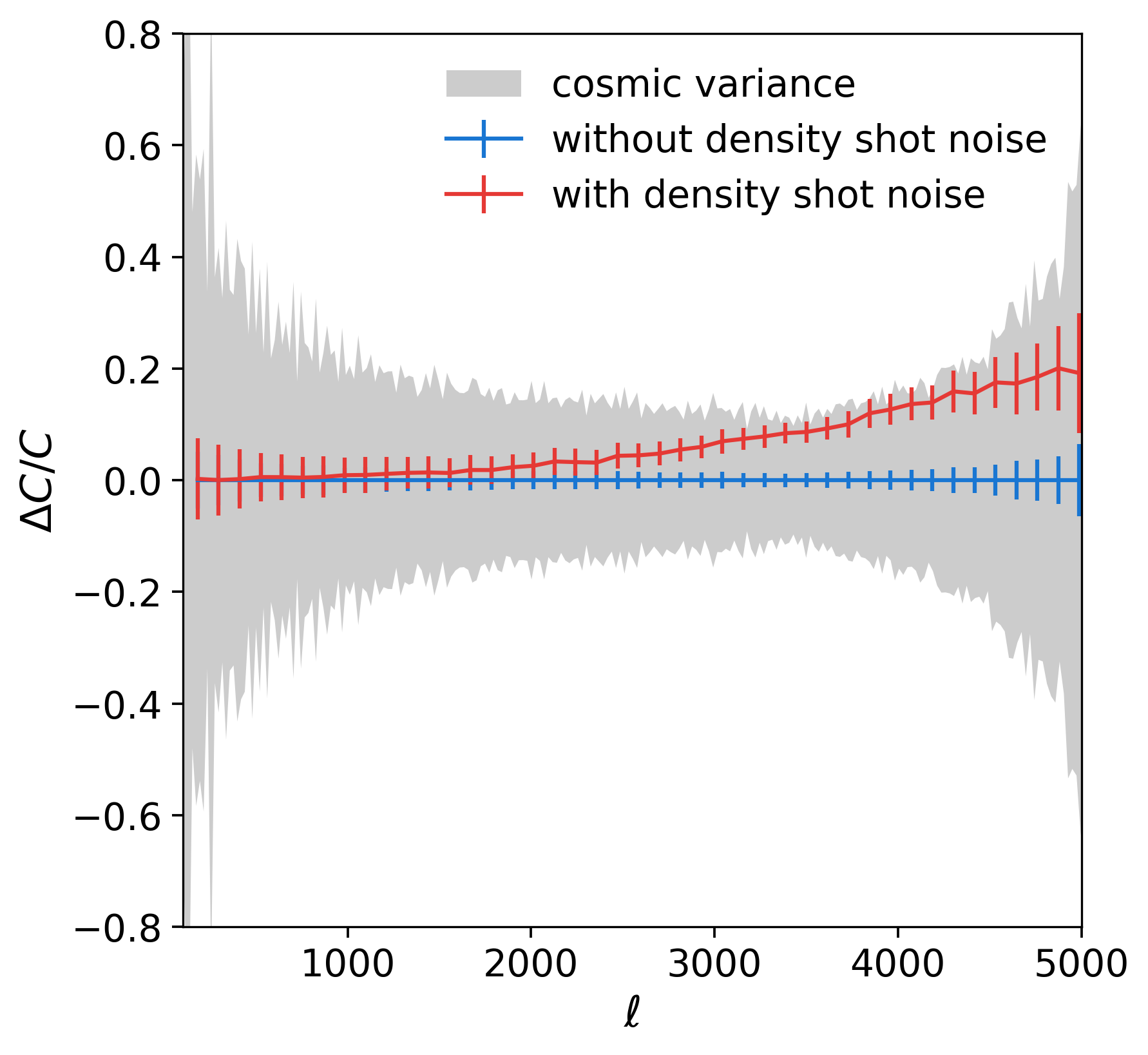}
    \caption{The impact of number density-induced shape noise on $C$ is presented in terms of the fractional difference relative to $\langle C(\kappatrue)\rangle$ where the number density-induced shape noise is removed (achieved by creating maps using $\rho_i$ and $\grms^2$ at ten times the fiducial value). The high density case is represented in blue, and its $1\sigma$ spread is shown in gray (as in \Fig\ref{fig:aliasing_cl}). $\mean{C(\kappatrue)}$ calculated with fiducial values for $\rho_i$ and $\grms$ is shown in red. The error bars denote the standard error of the mean.}
    \label{fig:density_cl}
\end{figure}

Finally, we observe another novel source of error that impacts FLI—the finite galaxy density can induce a shape noise-like error on the pixelized maps even when no shape noise is introduced. When we average the galaxy shapes within a pixel, the distribution of random galaxy positions samples the inhomogeneous shear field, leading to statistical fluctuation of the pixel value. Formally, for a pixel $p$ centered at $\bftheta_p$, we expect the galaxy's true shape to differ from $\gamma(\bftheta_p)$ by
\begin{align}
    \Delta\gamma \approx \frac{\partial\gamma}{\partial\theta}(\bftheta_p)\Delta\theta = \ell\gamma\Delta\theta\,.
\end{align}
This term behaves like a white noise that only appears in the autopower spectra of $\gammasys$. Therefore, we call this the number density-induced shape noise, or density shot noise in short. The impact of density shot noise is characterized in \Fig\ref{fig:density_cl}, where we see that it impacts the small scale autopower spectrum by about $20\%$ near the pixel resolution scales. We did not observe this noise in the crosspower spectra. 
\response{
To correct for this systematics, we first measure the power spectrum of this density-induced shape noise term ($C^\mathrm{dsn}_{ii}$) directly from mock catalogs for each redshift bin $i$ (as shown in \Fig\ref{fig:density_cl}). Then, when we generate convergence maps following \Eqs \ref{eq:chol} (for the Gaussian model) and \ref{eq:lognormal_cl} (for the log-normal model), we can simply change the cosmological power spectrum $C_{ii}$ via
\begin{equation}
    C_{ii} \rightarrow C_{ii} + C^\mathrm{dsn}_{ii} \,.
\end{equation}
In this way, the shear maps ($\gammasys_i$) generated by the forward model automatically include the density shot noise. The expression $(\gammaobs_i-\gammasys_i(\bfc,\bfx_i))^2/2N$ in \Eq\ref{eq:likelihood} still accurately describes the regular shape noise model and does not require additional modification.}

\section{Model misspecification}
\label{sec:modelmisspecs}

We have demonstrated that by accounting for various analysis systematics — including aliasing, boundary effects, mode coupling, and density-induced shape noise — we can obtain unbiased cosmological constraints from observed pixelized shear maps. A remaining modeling uncertainty is the field-level prior on the convergence field. It is anticipated that this uncertainty could potentially induce an absolute bias on the inferred parameters as well as affect the uncertainty quantification. In this section, we test both the Gaussian and the log-normal models using Gaussian, log-normal, and Takahashi mocks.

As we have shown above, a significant difference exists between the realistic observed shear maps and the maps generated by the forward model (which we shall call pipeline mocks). Previous works \cite{alsing_cosmological_2017,boruah_map-based_2022} have demonstrated that it is possible to use the forward model to analyze the pipeline mocks and correctly recover the cosmological parameters. Here, we follow  \citet{fiedorowicz_karmma_2022-1} and go one step further toward realism by using \miko\ to analyze the raw mocks themselves. A further feature of this analysis is that, to the best of our knowledge, this is the first study of its kind to extract cosmological parameters (here the $A_i$) using realistic mocks and taking into account the range of analysis systematics that must be considered for real data analysis. 

\subsection{Impact on cosmological parameters}
\subsubsection{The Gaussian model}
\begin{figure*}
    \centering
    \includegraphics[width=0.6\hsize]{ 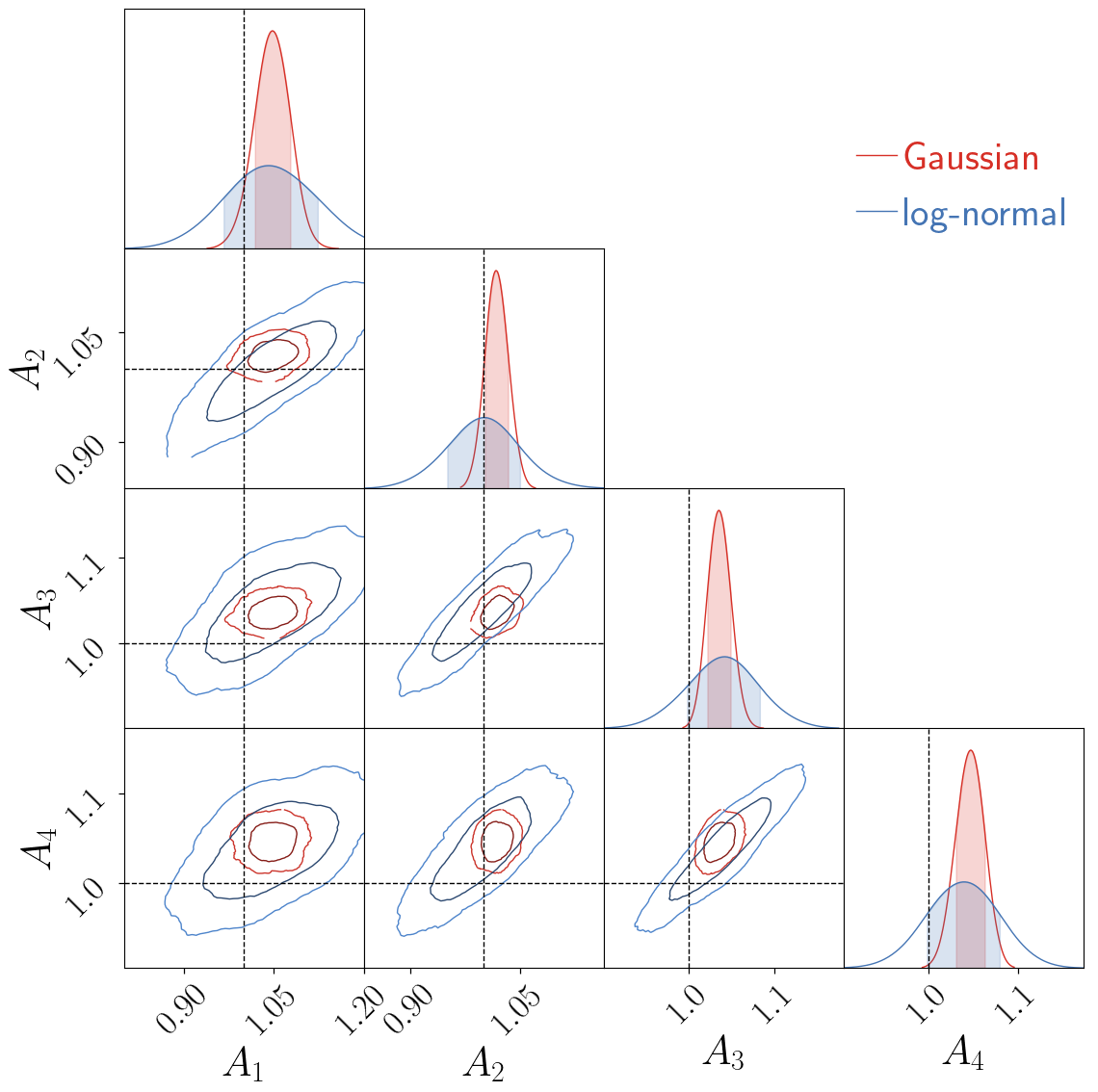}
    \caption{The joint posterior distribution of $A_i$ inferred from a $40\deg^2$ Takahashi patch using the Gaussian (red) and the log-normal (blue) models. The log-normal model employs an informative prior on $a_i$ that is jointly sampled (not shown here) with $A_i$. Compared to the Gaussian model, the log-normal model exhibits larger uncertainties and a stronger correlation between the $A_i$'s. The contours show the $1\sigma$ and $2\sigma$ level sets. 
    }
    \label{fig:corner1}
\end{figure*}

\begin{figure*}
    \centering
    \includegraphics[width=0.9\hsize]{ 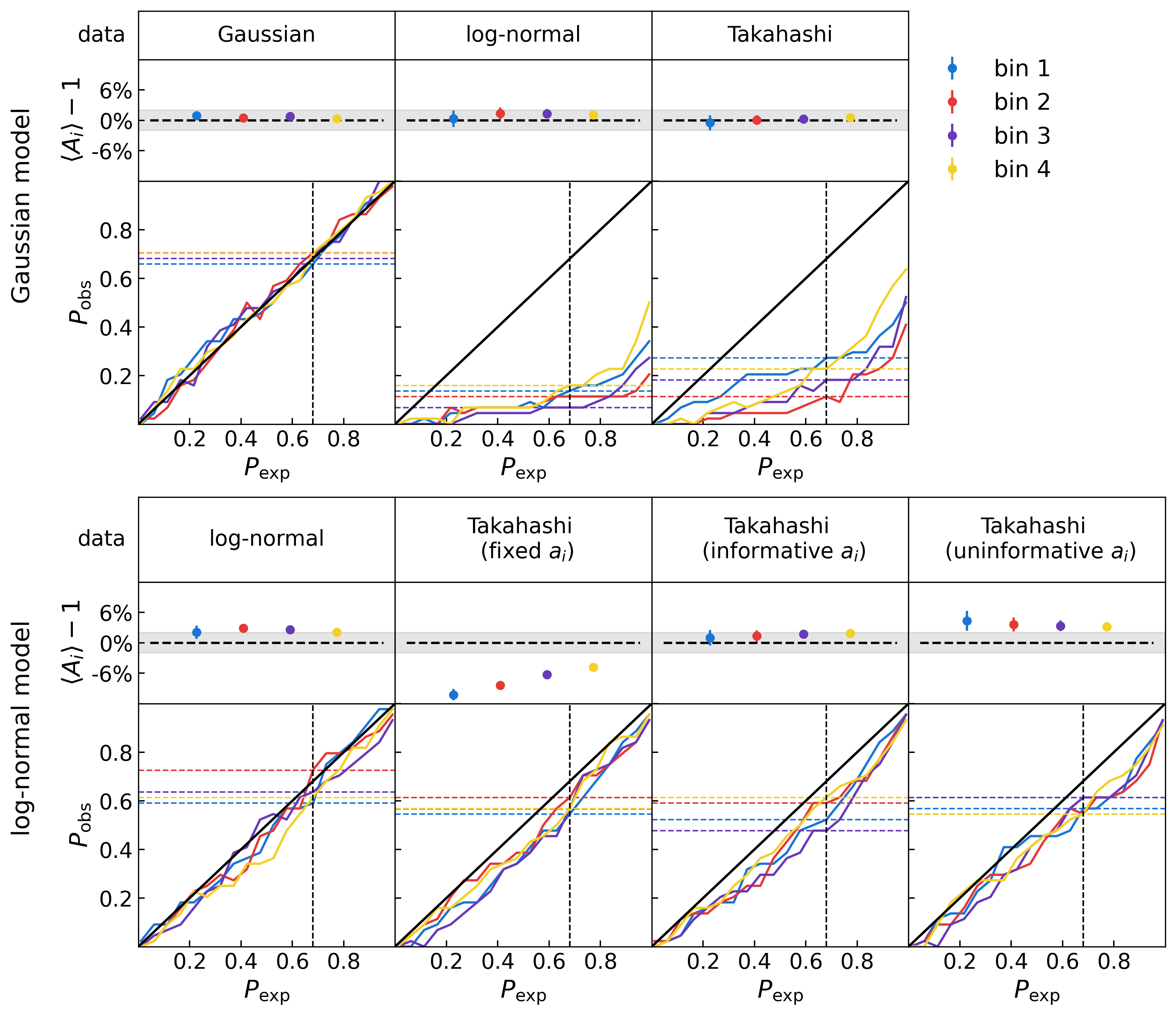}
    \caption{The top panel displays the results of the Gaussian model applied to Gaussian, log-normal, and Takahashi mocks. Within this panel, the first row indicates the absolute bias of $A_i$, averaged over $80$ independent HSC-like experiments. The gray area denotes the systematic error control target of $2\%$. The second row presents the P-P plot for $A_i$ (with the mean bias subtracted), where the x-axis denotes the expected probability ($P_\mathrm{exp}$) that a credible interval contains the true $A_i$, and the y-axis denotes the observed probability ($P_\mathrm{obs}$). The vertical dashed line indicates $P_\mathrm{obs} = 68\%$, and the colored horizontal lines show where the P-P curves intersect with this line. The Gaussian model demonstrates no absolute bias irrespective of model misspecification but exhibits overconfidence in the error bars when applied to realistic data. The bottom panel illustrates the absolute bias and the P-P plot for $A_i$ using the log-normal model, applied to log-normal and Takahashi mocks. For the Takahashi mocks, three scenarios are tested with the log-normal model concerning the shift parameters: 1) fixed at values measured from noiseless true mocks (fixed $a_i$), 2) allowed to vary within a narrow and informative range, and 3) allowed to vary within a broad and uninformative range. The log-normal model generally yields accurate uncertainties, yet the absolute bias is highly sensitive to the choice of $a_i$.}
    \label{fig:A_both_priors}
\end{figure*}

First, we run the Gaussian model on all three sets of mocks. For each model-data pair, we perform $80$ HSC-like experiments on independent patches, utilizing analysis systematics models recommended by \Sec\ref{sec:analysissys}. For each experiment, we obtain the joint posterior distribution of $A_i$ along with the map parameters. For instance, the constraint on $A_i$ for a single Takahashi patch is depicted as the red contour in \Fig\ref{fig:corner1}. The $A_i$'s are uncorrelated in the Gaussian model, consistent with the findings in \citet{zhou_field-level_2023}. There is considerable cosmic variance within a $40\deg^2$ patch. To determine whether the pipeline has an absolute bias on $A_i$, we compute $\mean{A_i} - 1$ over independent experiments and present the results in the top panel of \Fig\ref{fig:A_both_priors}. The top part of the top panel shows that the Gaussian model does not exhibit absolute bias for all simulated datasets, regardless of whether the data is simulated with a Gaussian model or not. 

To assess the impact of model misspecification on our uncertainty estimates, let us consider the one-dimensional marginal posterior distribution of a particular $A_i$. Ideally, $x\%$ (the observed probability) of all the experiments should find $A_i=1$ contained within the $x\%$ (the expected probability) credible interval. The observed probability as a function of the expected probability is called the P-P plot. It is shown in the lower part of the top panel of \Fig\ref{fig:A_both_priors}, after adjusting for mean absolute bias to concentrate on the error bars. Perfect uncertainty prediction would align the P-P curves with the line $y=x$. This alignment occurs only without model misspecification. For instance, applying the Gaussian model to log-normal or Takahashi mocks results in overly optimistic uncertainty estimates for all $A_i$ values, particularly for log-normal mocks where the truth is included within the one standard deviation interval less than $20\%$ of the time. This is consistent with the observation of \citet{boruah_map-based_2022}, although the extent to which the Gaussian model underestimates the uncertainty is much more severe in our case. The bottom line is that assuming a Gaussian prior leads to an unbiased estimate of the cosmological parameters but a {\it biased} estimate of the errors on those parameters.

\subsubsection{The ``unreasonable'' effectiveness of the Gaussian priors}
\begin{figure*}
    \centering
    \includegraphics[width=1\hsize]{ 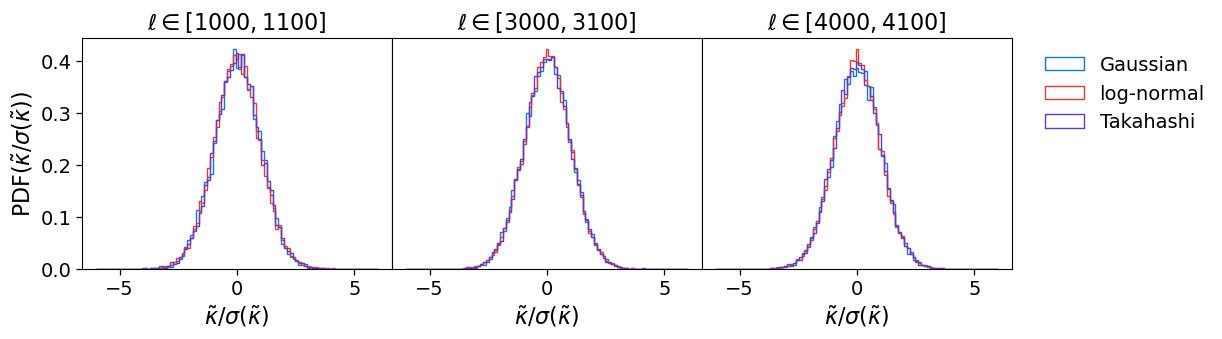}
    \caption{Standardized one-point PDF of $\fs\kappa(\ell)$ for Gaussian, log-normal, and Takahashi mocks at three distinct $\ell$ values. Across all mocks and $\ell$ values, the distribution of $\fs\kappa(\ell)$ remains consistently Gaussian, which elucidates the Gaussian model's ability to infer cosmological parameters without absolute bias, notwithstanding model misspecification.}
\label{fig:pdf_fs}
\end{figure*}

Why does the Gaussian map prior not induce any absolute bias on $A_i$ when it is applied to clearly non-Gaussian fields? The answer lies in the PDF of $\fs\kappa$ in Fourier space. In \Fig\ref{fig:pdf_fs}, we plot the PDFs of $\fskappatrue$ for Gaussian, log-normal, and Takahashi mocks at three different $\ell$ values. Interestingly, for all mocks and across all ranges of $\ell$, the PDF of $\fskappatrue$ is consistent with a Gaussian distribution. Similar phenomena have previously been observed in three-dimensional simulations, for example, by \citet{matsubara_statistics_2007} and \citet{qin_numerical_2022}. 
Motivated by these observations, we can express the PDFs of non-Gaussian ($\prior^\mathrm{NG}$) and Gaussian ($\prior^\mathrm{G}$) convergence fields as follows (ignoring aliasing and other analysis systematics for simplicity)
\begin{equation}
    \prior^\mathrm{NG}(\fs\kappa) = \prior^\mathrm{G}(\fs\kappa) \left[1+\delta(\fs\kappa)\right] \,.
\end{equation}
In the absence of model misspecification, the likelihood analyses should yield unbiased $A^\text{NG}_i$, where
\begin{align}
    \nonumber
    & \langle A^\text{NG}_i \rangle \\
    & = \int d\gammaobs \int d\fs\kappa ~ A_i ~ \post(\gammaobs|\fs\kappa,A_i) ~ \prior^\mathrm{NG}(\fs\kappa) \\
    & = \int d\gammaobs \int d\fs\kappa ~ A_i ~ \post(\gammaobs|\fs\kappa,A_i) ~ \prior^\mathrm{G}(\fs\kappa) \left[1+\delta(\fs\kappa)\right] \\
    & = \langle A^\text{G}_i \rangle + \text{correction}\left[\delta\right]
\end{align}
Therefore, as long as the perturbation $\delta$ is small, $\langle A^\text{G}_i \rangle = \langle A^\text{NG}_i \rangle$ is unbiased. 

\begin{figure}
    \includegraphics[width=0.9\hsize]{ 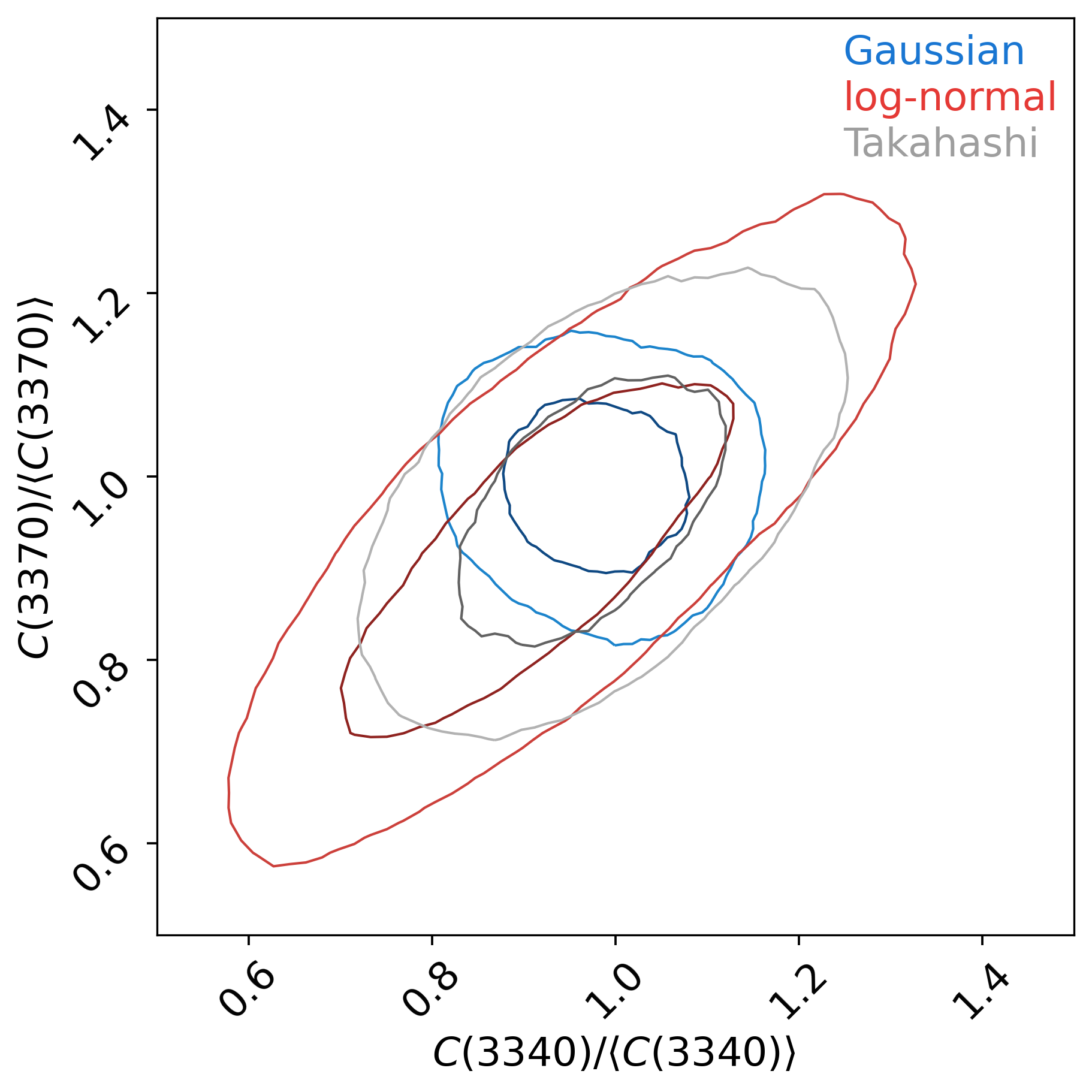}
    \caption{The correlation between $C(\ell_1)/\langle C(\ell_1) \rangle$ and $C(\ell_2)/\langle C(\ell_2) \rangle$ for Gaussian (blue), log-normal (red), and Takahashi (gray) mocks. In this example, we have chosen $\ell_1 = 3340$ and $\ell_2 = 3370$, and $C$ to be the autopower spectrum of the first redshift bin. The contours show the $0.3$ and $0.8\sigma$ level sets. The Gaussian model assumes no mode correlations. The log-normal model assumes more mode correlation than the realistic mocks.} 
    \label{fig:cl_corr}
\end{figure}

\begin{figure}
    \includegraphics[width=1\hsize]{ 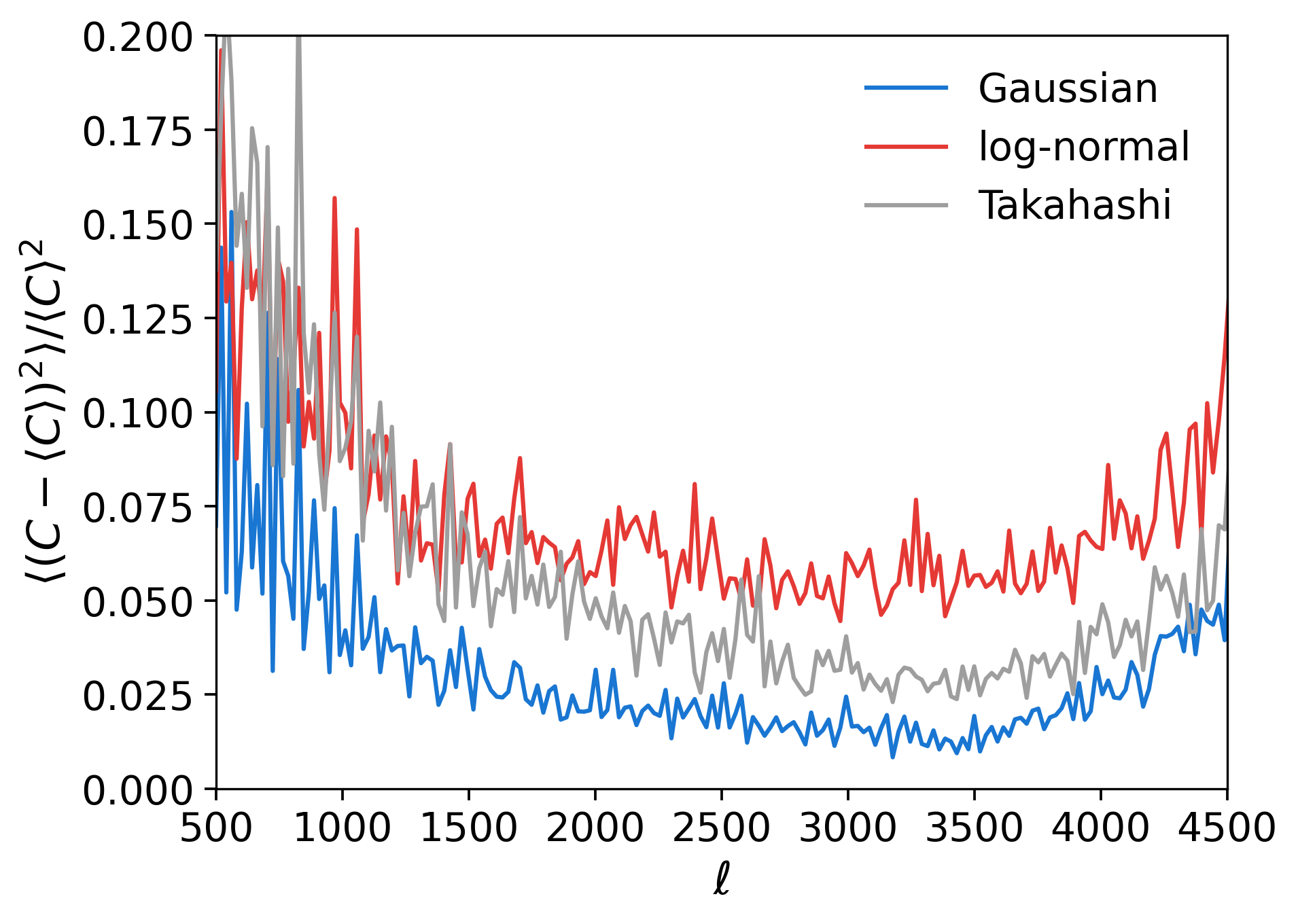}
    \caption{Normalized variance of the power spectrum for Gaussian (blue), log-normal (red), and Takahashi (gray) mocks, with $C$ representing the autopower spectrum of the first redshift bin. Notably, the Gaussian mocks exhibit smaller cosmic variance relative to the Takahashi mocks, whereas the log-normal mocks demonstrate greater cosmic variance in comparison.}
    \label{fig:4pt}
\end{figure}

The second question pertains to why the Gaussian model fails to predict the correct uncertainty of $A_i$. To address this, we first examine the correlation between $C(\ell_1)$ and $C(\ell_2)$ across the Gaussian model, the log-normal model, and the Takahashi mocks, as illustrated in \Fig\ref{fig:cl_corr}. As expected, the Gaussian model assumes there is no correlation between the power of two different $\ell$ modes. Nonetheless, correlations are evident in both the log-normal simulations and the realistic Takahashi mocks. Consequently, the Gaussian model presupposes that each $\ell$ mode contributes independently, while in actuality, the different $\ell$ modes are correlated and thus provide less information than assumed. As a result, the Gaussian model reports a smaller error bar on the inferred power spectrum amplitude $A_i$, an error analogous to assuming a larger $f_\mathrm{sky}$ than what is true in reality. Another perspective on this issue is provided by examining the cosmic variance predicted by the Gaussian, log-normal, and Takahashi maps, as shown in \Fig\ref{fig:4pt}. The Gaussian model predicts smaller cosmic variance than the other two models across all scales, which then translates to the narrower uncertainty bounds on the estimated power spectrum amplitude $A_i$.


\subsubsection{The log-normal model}
\label{sec:lognormal_result}
The log-normal model depends on both $A_i$ and the shift parameter $a_i$. We first run the log-normal model on the $44$ log-normal mocks fixing $a_i$ at the truth. The pipeline recovers the correct $A_i$ with an absolute bias within $2\%$\footnote{The slight positive bias is due to the deficit of large-scale powers discussed in \Sec\ref{sec:modecoup}. This error can be alleviated by using larger patches.}). We also report excellent uncertainty quantification. This result is shown in the bottom panel of \Fig\ref{fig:A_both_priors} (column 1).

Model misspecification also arises when applying the log-normal model to the Takahashi mocks. Specifically, since the Takahashi mocks do not adhere to a strict log-normal distribution, as shown in \Fig\ref{fig:field_example}, a definitive ground truth for $a_i$ is absent. Previous studies have approached $a_i$ in various ways: for instance, \citet{alsing_cosmological_2017} treated $a_i$ as fixed parameters, whereas \citet{boruah_map-based_2022} and \citet{fiedorowicz_karmma_2022-1} employed perturbation theory for approximation. As we will show, the modeling of $a_i$ is arguably the primary source of bias with the log-normal model and warrants careful consideration.

In our first approach, we compute $a_i^\text{fit}$ by fitting the one-point PDF to the noiseless Takahashi mocks using \Eq\ref{eqn:shift_skewness}. We then perform inference on the Takahashi mocks assuming $a_i = a_i^\text{fit}$. This method approximates a perfect perturbation theory prediction for $a_i$. The result is presented in the bottom panel of \Fig\ref{fig:A_both_priors} (column 2). Surprisingly, even when using the optimally fitted $a_i^\text{fit}$, the log-normal model infers $A_i$ with a significant negative bias. This bias is most pronounced, reaching up to $10\%$, in the lowest redshift bins where the convergence field exhibits the highest non-Gaussianity. Consequently, we deduce that merely employing the log-normal shift parameter that most closely fits the PDF of the actual convergence field is not only insufficient but also incorrect for acquiring unbiased constraints on cosmological parameters.

\begin{figure*}
    \centering
    \includegraphics[width=0.6\hsize]{ 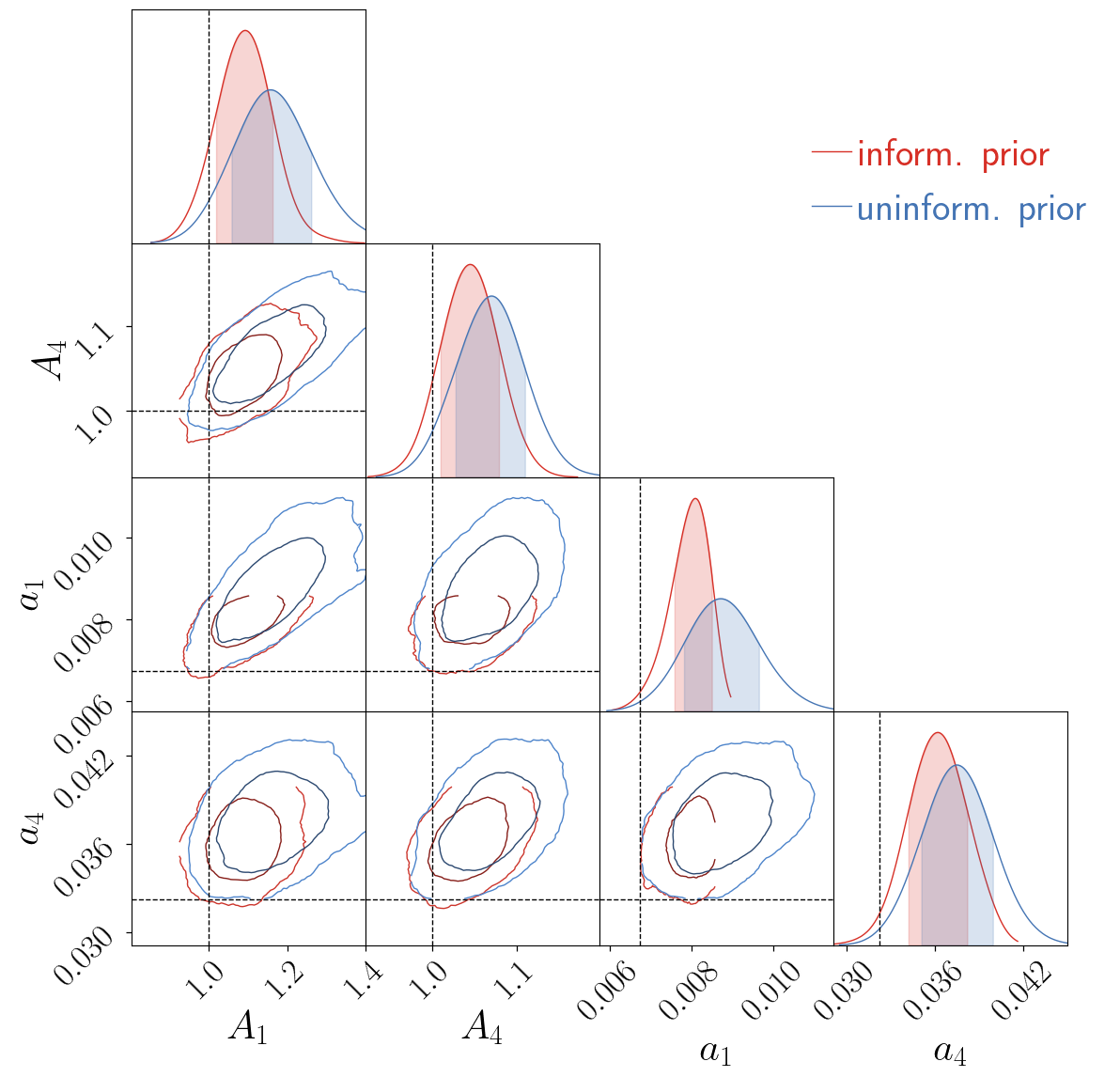}
    \caption{
    The joint posterior distribution of $A_i$ and $a_i$, as inferred from a Takahashi patch using the log-normal model. We compare an informative prior (red) and an uninformative prior (blue) on $a_i$. For simplicity, only the first and the last redshift bins are shown. The dashed lines across $A_i$ represent $A_i=1$, and the dashed lines across $a_i$ represent $a_i^\text{fit}$—the best-fit values from noiseless data. Three important conclusions are 1) $A_i$ and $a_i$ are positively correlated, 2) the log-normal model prefers $a_i$ much larger than $a_i^\text{fit}$, and 3) the informative prior truncates the distribution of $a_i$. Therefore, when the prior is relaxed, $A_i$ shifts to higher values. The contours show the $1\sigma$ and $2\sigma$ level sets. 
    }
    \label{fig:corner2}
\end{figure*}

In our second approach, we adopt an agnostic stance on $a_i$, treating it as a variable during sampling. Observing that the likelihood analysis typically favors $a_i > a_i^\text{fit}$, we implement a flat prior on $a_i$ in the range $\left[ 0.9 a^\text{fit}_i, 1.3 a^\text{fit}_i \right]$. The joint posterior of $A_i$, as depicted in \Fig\ref{fig:corner1}, reveals that the uncertainties reported by the log-normal model are larger—yet appear more realistic—than those by the Gaussian model, with $A_i$ exhibiting positive correlation. Furthermore, the joint posterior of $A_i$ and $a_i$ (\Fig\ref{fig:corner2}) indicates a positive correlation between them, with the maximum likelihood estimate for $a_i$ substantially exceeding $a_i^\text{fit}$. This is consistent with the previous observation that fixing $a_i$ to $a_i^\text{fit}$ induces a negative bias on $A_i$. 
Nonetheless, \Fig\ref{fig:corner2} also illustrates that the prior range for $a_i$ truncates the posterior distribution, hence we refer to this as the informative prior model. Through $44$ independent experiments, we find that imposing an informative prior on $a_i$ can significantly reduce the absolute bias on $A_i$ to within $2\%$ while maintaining accurate uncertainty estimation for $A_i$. 
The P-P curves in the bottom panel of \Fig\ref{fig:A_both_priors} quantify the extent to which the uncertainty estimation is accurate when the log-normal prior is used.

Lastly, we impose a uniform and uninformative prior on $a_i$ over the positive reals. We explicitly checks that the prior does not truncate the posterior distribution. An example is shown in \Fig\ref{fig:corner2}. As the positive correlation between $a_i$ and $A_i$ suggest, this model creates a positive absolute bias on $A_i$ (bottom panel of \Fig\ref{fig:A_both_priors}, column 4). The bias is most severe in the lowest redshift bin at the $4\%$ level. 

To summarize, the three choices for $a_i$'s prior suggest that, in order to obtain unbiased cosmological constraints on real data using the log-normal model, we can neither use $a_i$ directly measured from the noiseless mocks nor treat it as a variable that is completely free. Otherwise, although we obtain good uncertainty quantification, we risk significant bias in cosmological parameters. The existence of the absolute bias is not too surprising given that the log-normal also misspecifies the field-level priors.

\subsection{Map reconstructions}
\label{sec:map_result}
\begin{figure*}
    \centering
    \includegraphics[width=1\hsize]{ 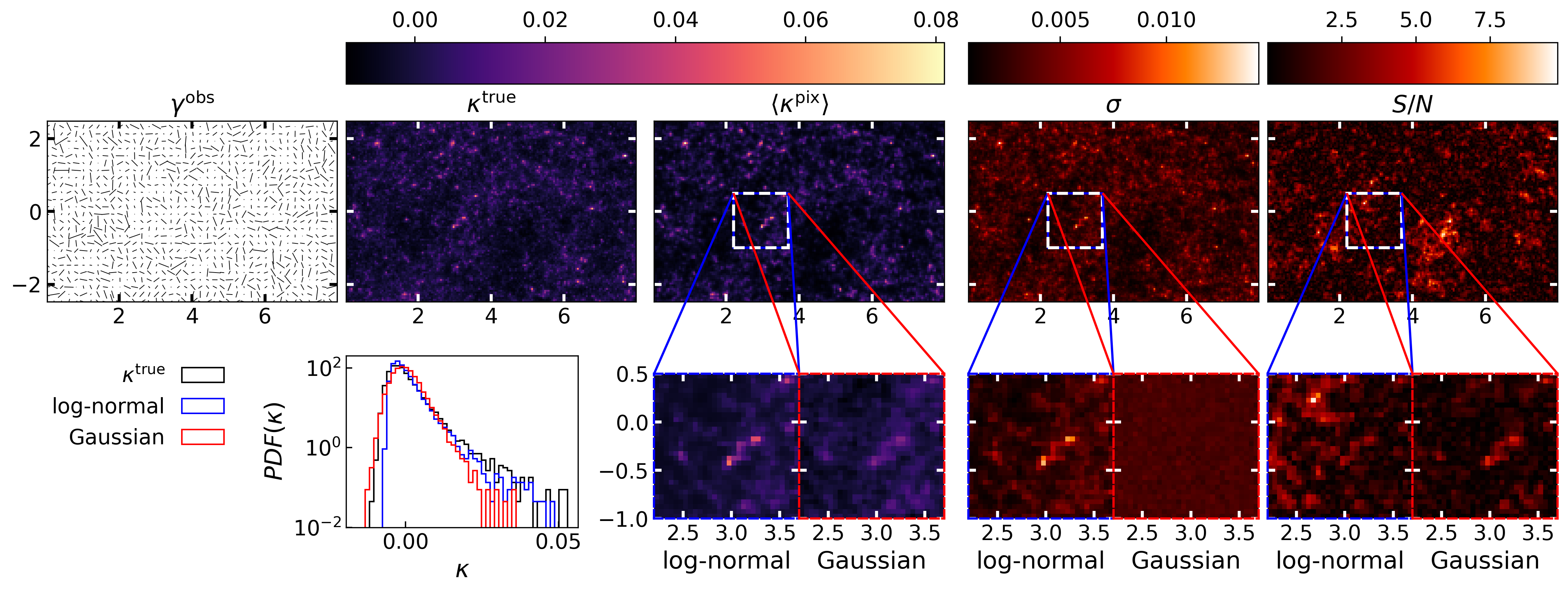}
    \caption{Field-level reconstruction for the first redshift bin of a single Takahashi patch using the log-normal (with informative prior on $a_i$) and the Gaussian models. The first row, from left to right, are the observed noisy shear map $\gammaobs$, the true convergence map $\kappatrue$, the mean posterior map $\mean{\kappasys}$ (log-normal model), the standard deviation map $\sigma$ (log-normal model), and the S/N map (log-normal model). In the second row, the leftmost plot compares the histograms of $\kappatrue$ with those of $\mean{\kappasys}$ for both Gaussian and log-normal models. The three inset plots zoom in on $1.5\deg \times 1.5\deg$ region on the patch, contrasting the log-normal results (left) with the Gaussian results (right).}
\label{fig:field}
\end{figure*}

\miko~also provides the joint posterior distribution of all the map pixels—$(\kappasys_i)_s$ (the bottom diamond in the flowchart \Fig\ref{fig:flowchart}), where $s$ denotes the sample index. For simplicity, let us focus specifically on one experiment on the Takahashi mock, and on the reconstruction of the first redshift bin convergence field. The results for both the Gaussian and log-normal (informative prior) models are summarized in \Fig\ref{fig:field}.

Let us start with the first row. The first and second columns depict the observed noisy shear field, $\gammaobs$, and the true convergence, $\kappatrue$, respectively. The final three columns pertain exclusively to the log-normal model, illustrating the mean posterior map $\mean{\kappasys}$, the standard deviation map $\sigma$, and the signal-to-noise ratio map (S/N). For these last three maps, we zoom in on a $1.5\deg \times 1.5\deg$ patch in the inset plots below. In these insets, we also compare the log-normal results with those of the Gaussian model.

Theoretically, we do not expect $\mean{\kappasys}$ to match $\kappasys$ exactly. Instead, $\mean{\kappasys}$ is \emph{roughly} analogous to a Wiener-filtered version of $\kappasys$. This filtering smooths the map on small scales where shot noise dominates (see \citet{zhou_field-level_2023} for a detailed discussion on the properties of the posterior maps and the maps' power spectra). Both the Gaussian and the log-normal model recover the large-scale features of the true convergence maps. The log-normal model, however, resolves the density peaks much better (insets in the second row). The adjacent histogram plot further supports this claim - the log-normal model's PDF$(\mean{\kappasys})$ agrees with PDF$(\kappatrue)$ up to the highest $\kappa$ values. In contrast, the Gaussian model's PDF$(\mean{\kappasys})$ decays much faster. On the other hand, the log-normal model's PDF$(\mean{\kappasys})$ shows a hard cutoff in the low $\kappa$ region. The Gaussian model captures the voids much better in comparison. This aligns with the findings of \citet{fiedorowicz_karmma_2022-1}, who reported that the log-normal model is less successful in reconstructing the correct void count distribution at higher resolutions.

The log-normal and the Gaussian model also differ in their prediction of the pixel-level uncertainty. (We show the $1\sigma$ uncertainty for each pixel in the fourth column and the comparison between the log-normal model and the Gaussian model in the inset plots below.) The variance predicted by the log-normal model shows a positive correlation with the absolute pixel value, whereas the Gaussian model predicts uniform uncertainty across all pixels. In the rightmost column, we display the signal-to-noise ratio (S/N). Interestingly, both models predict identical S/N for the peaks, yet the log-normal model yields a higher S/N for the voids due to its nonuniform uncertainty predictions.

\begin{figure}
    \centering
    \includegraphics[width=1\hsize]{ 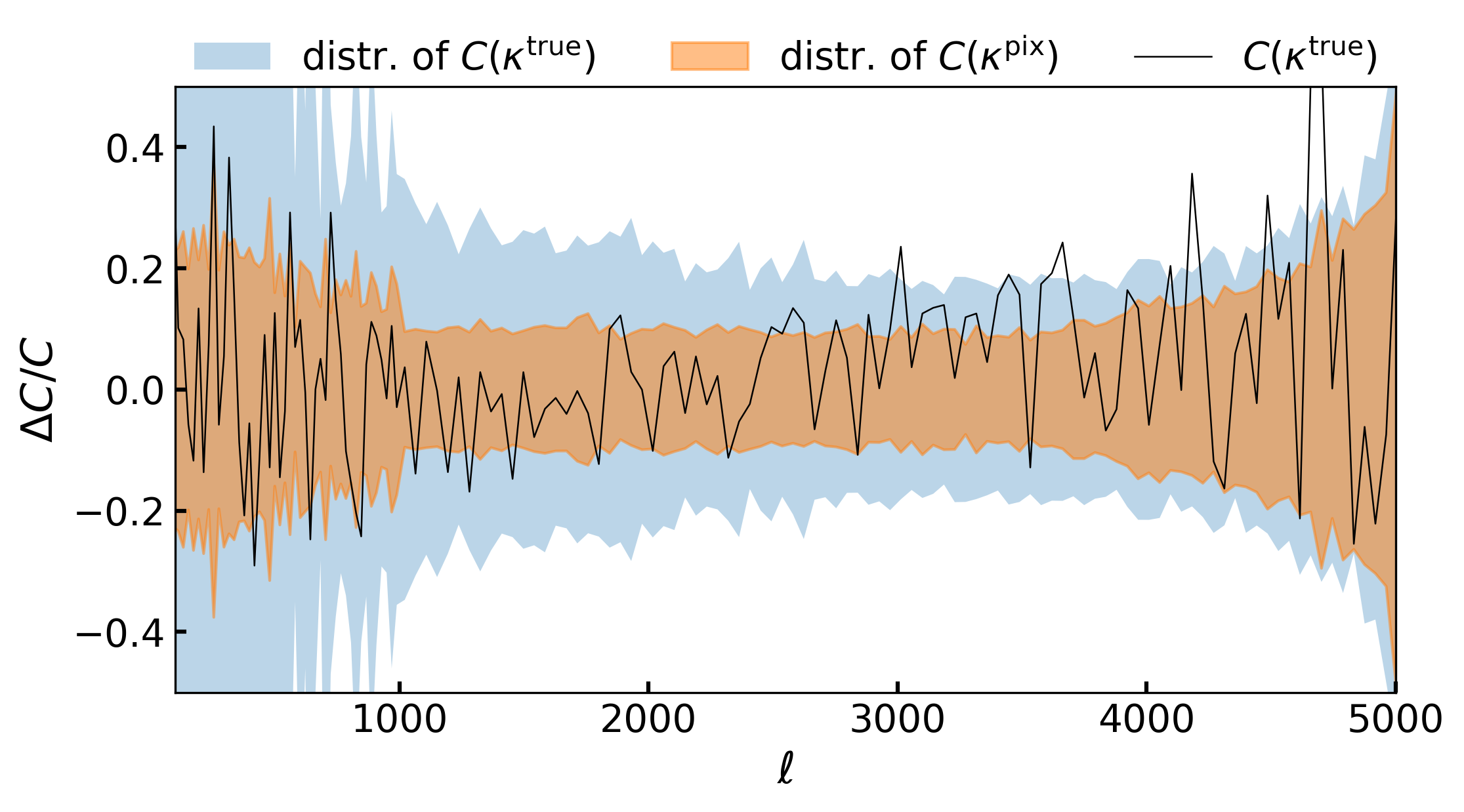}
    \caption{Power spectra is presented in terms of the fractional difference relative to $\mean{C(\kappasys)}$, where $\kappasys$ are map samples inferred from a single Takahashi patch using the log-normal model with an informative prior on $a_i$. The power spectrum distribution of the sampled maps, $C(\kappasys)$, is depicted in orange, while the distribution of the mock fields, $C(\kappatrue)$, is shown in blue to indicate cosmic variance. The power spectrum of the true noiseless field being inferred is shown in black. In this context, $C$ specifies the autopower spectrum of the first redshift bin. The posterior samples are in agreement with $\kappatrue$ at the two-point level within the bounds of uncertainty.}
    \label{fig:lnontakacl}
\end{figure}

For each $\kappasys$ sample, we can compute its power spectrum $C(\kappasys)$. As proved in \citet{zhou_field-level_2023}, in the absence of model misspecification, the distribution of $C(\kappasys)$ is expected to follow that of $C(\kappatrue)$. The distribution of $C(\kappasys)$ for the log-normal model with informative prior on $a_i$ is depicted in \Fig\ref{fig:lnontakacl}, where we indeed confirm that $C(\kappasys)$ recovers $C(\kappatrue)$ within uncertainty. 

\section{summary}
\label{sec:summary}
We have presented a percent-level accurate field-level cosmological analysis pipeline, \miko{}, that infers the joint posterior distribution of the cosmological parameters and the tomographic convergence fields. We identified several analysis systematics such as aliasing, boundary effect, mode-coupling, and density-induced shape noise, many of which have not been captured or modeled in previous studies. We measured their impact on cosmological parameter constraints and proposed effective mitigation methods that control the systematic uncertainty down to the $2\%$ level. We then explore the problem of model misspecification (using wrong field-level priors) by testing the Gaussian and log-normal forward models on three sets of mock data (Gaussian, log-normal, and ray-tracing simulations). The Gaussian model shows no absolute bias in the $A_i$ (the amplitude of the power spectrum in the $i^{th}$ redshift bin) even with model misspecification, but yields overconfident uncertainty on cosmological parameters. The log-normal provides accurate uncertainty quantification, but is sensitive to the choice of the shift parameters, $a_i$. In particular, the shift parameters fitted to the simulations can incur $10\%$ absolute bias on power spectrum amplitudes. After accounting for analysis systematics (and calibrating $a_i$'s prior for the log-normal model), both models recover $A_i$ with an absolute bias less than $2\%$. Our result demonstrates the accuracy and unbiased nature of this field-level cosmological analysis framework, a prerequisite for it to provide meaningful, tight cosmological parameter constraints. 

The model misspecification bias identified in this work carries important practical implications for future research. Given that the log-normal distribution results in prior-dependent cosmological constraints, it is crucial to calibrate $\prior(a_i)$ before analyzing real data. We have demonstrated the feasibility of constructing $\prior(a_i)$ to infer unbiased cosmology from mocks. Nonetheless, there is no assurance that the same $\prior(a_i)$ will work on real data distributions. Future studies should examine the robustness of $\prior(a_i)$ across different simulations or investigate alternative one-point distributions that both reflect the non-Gaussianity of the field and remain unaffected by priors on nuisance parameters. Given the similar cosmological constraining powers derived from both the Gaussian and the log-normal models, it currently seems more practical to utilize the Gaussian model and calibrate the uncertainty estimates based on independent mock datasets.

Many analysis systematics identified in this work are applicable to general field-level inference problems involving pixelized maps. For instance, we can employ the same aliasing correction when modeling the observed temperature and polarization maps CMB at the field level. Otherwise, we risk a positive absolute bias in the inferred amplitude of the power spectrum. The error induced by model misspecification is also not weak lensing-specific. 

In the next step, we will implement galaxy intrinsic alignment \cite{troxel_intrinsic_2015}, redshift error \cite{HSC3_photoz_Rau2022}, and point spread function \cite{HSC3_PSF} systematics into \miko{} and apply to the HSC Year 3 data. Implementing efficient algorithms to incorporate correlated pixel noise is also important \cite{millea_optimal_2021}. Another direction is to leverage emulator-based models to constrain more general cosmological parameters \cite{nishimichi_dark_2019,kobayashi_full-shape_2022,miyatake_hyper_2023}.

\section*{Acknowledgment}
We thank Masahiro Takada, Fabian Schmidt, and Bhuvnesh Jain on detailed and insightful comments on the draft of this paper. A.Z. thanks Chirag Modi for discussion on sampling, Benjamin Horowitz for discussion on the log-normal map prior, and Yin Li for discussion on the FFTLog Hankel transform in early stages of the project. A.Z. also expresses his gratitude to the Yukawa Institute for Theoretical Physics at Kyoto University for hosting the YITP workshop YITP-W-23-02 on "Future Science with CMB x LSS". The discussions during the workshop were instrumental in completing this work. X.L. and R.M. were supported by a grant from the Simons Foundation (Simons Investigator in Astrophysics, Award ID 620789). This work was supported by NSF Award Number 2020295.

\bibliographystyle{mnras}
\bibliography{references}

\appendix
\begin{figure}
    \centering
    \includegraphics[width=1\hsize]{ 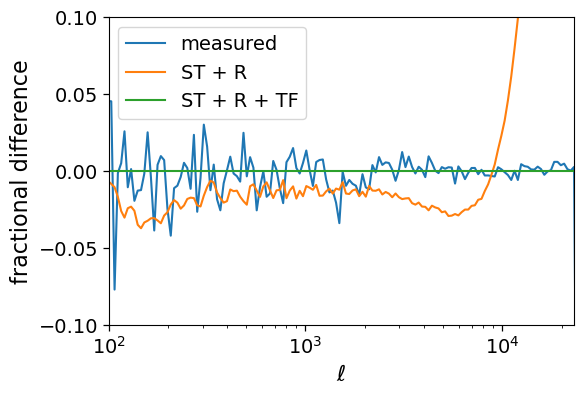}
    \caption{Power spectra of the first redshift bin in terms of the fractional difference relative to the adopted theoretical $C$ for the Takahashi mocks (green). The mean of $C$, as measured on Takahashi mocks, is depicted in blue. The theoretical $C$, corrected only for shell thickness and resolution, is shown in orange. This correction is found to be insufficient for meeting our accuracy requirements. Abbreviations used: ST for shell-thickness correction, R for resolution correction, and TF for transfer function correction.}
    \label{fig:cl_caliberation}
\end{figure}

\section{Caliberation of the Takahashi simulations' power spectra}
\label{app:cl_calibration}
The convergence and shear maps calculated from the Takahashi simulations have a scale-dependent and redshift-dependent bias on the two-point level when compared to the theoretical prediction (\Eq\ref{eqn:2pt}). When we analyze the Takahashi simulations, we need to correct our theoretical model to match the simulation. At low redshift, the dominating systematics is the discretization of the lens plane. During ray tracing, the simulation assumes the light rays are deflected by lens planes of finite thickness $\Delta^\text{ST}\!=\!150h^{-1}\ \Mpc$. We therefore introduce a correction in \Eq\ref{eqn:2pt} by convolving the $P_\delta(k;\chi)$ with a window function \cite{takahashi_full-sky_2017}
\begin{equation}
    P_\delta(k;\chi) \longrightarrow \Delta^\text{ST} \int \frac{dk_r}{2\pi} P_\delta(k;\chi) \sinc^2\left(\frac{k_r \Delta^\text{ST}}{2}\right)
\end{equation}
where $k_r$ is the radial wave vector. At high-$\ell$, the bias is mostly due to the pixelization effect. This can be corrected by \cite{takahashi_full-sky_2017}
\begin{equation}
    C_{ij}(\ell) \longrightarrow \frac{C_{ij}(\ell)}{1+(\ell/ (1.6\ \nside))^2}
\end{equation}
These two corrections reduce the bias below $5\%$ for $\ell<10000$ as shown in orange in \Fig\ref{fig:cl_caliberation}. Finally, we fit a transfer function so that the bias falls below $1\%$ for $\ell<24000$ for all redshift bins (shown in green).

\end{document}